\documentclass[10pt]{article}
\usepackage{epsfig,amssymb,amsmath,psfrag}
\usepackage[T1]{fontenc}
\usepackage{psfrag}
\usepackage{geometry}
\numberwithin{equation}{section}

\catcode`\@=11
\def \be {\begin{equation}}
\def \ee {\end{equation}}
\def \ba {\begin{eqnarray}}
\def \ea {\end{eqnarray}}
\def \baa {\begin{eqnarray*}}
\def \eaa {\end{eqnarray*}}
\def \bbib {\begin {thebibliography} }
\def \ebib {\end{thebibliography}}
\def \lab #1 {\label{#1}}

\def \matrix #1 {\left(\begin{array}{cc} #1 \end{array}\right)}

\def \Im {\mathop{\rm Im}\nolimits}

\def \e {\mathop{\rm e}\nolimits}

\newcommand{\as}{\ifmmode\alpha_{\rm s}\else{$\alpha_{\rm s}$}\fi}
\newcommand{\asbar}{\ifmmode\bar{\alpha}_{\rm s}\else{$\bar{\alpha}_{\rm s}$}\fi}

\newcommand{\ft}[2]{{\textstyle\frac{#1}{#2}}}

\font\cmss=cmss12 

\def\IZ{\relax{\hbox{\cmss Z\kern-.4em Z}}}
\def\IR{{\hbox{{\rm I}\kern-.2em\hbox{\rm R}}}}

\def\IP{{\hbox{{\rm I}\kern-.2em\hbox{\rm P}}}}
\def\II{\hbox{{1}\kern-.25em\hbox{l}}}

\begin{document}
\addtolength{\baselineskip}{1.3pt}
\bibliographystyle{unsrt}
\title{$R$-current six-point correlators in $AdS_5$ Supergravity}
\author{J.~Bartels$^{1}$, J.~Kotanski$^{1}$, A.-M.~Mischler$^{1}$,
V.~Schomerus$^{2}$
\bigskip\\
{\it $^1$~II. Institute Theoretical Physics, Hamburg University, Germany}\\
{\it $^2$~DESY Theory Group, Hamburg, Germany}\\
}
\maketitle
\begin{abstract}
\noindent Within the conjectured duality between ${\cal N}=4$
super Yang-Mills and Anti-deSitter string theory, the BFKL Pomeron
of the gauge theory corresponds to the graviton mode of the dual
string. As a first step towards analyzing multigraviton exchange,
we investigate $R$-current six-point functions within the
supergravity approximation. We compute the analogue of diffractive
scattering, and we analyze the triple Regge limit. In the
supergravity approximation the triple graviton vertex is found to
vanish.
\end{abstract}

\vspace{-9cm}
\begin{flushright}
{\small DESY--09--217}\\
\end{flushright}
\vspace{8.5cm}

\begin{flushleft}
\end{flushleft}

\section{Introduction}

Since many years, the high energy behavior of scattering
amplitudes in quantum field theory has attracted interest, and
extensive calculations have been performed in order to understand
the structure well beyond leading orders of perturbation theory.
In this context, a special role is played by the Regge limit which
is closely connected with unitarity of the theory.

The AdS/CFT
correspondence~\cite{Polyakov:1980ca,Maldacena:1997re,Witten:1998qj,Gubser:1998bc}
has raised new hopes to determine the high energy behavior to all
orders of the 't Hooft coupling $\lambda$, including the strong
coupling region, at least for those gauge theories which possess a
dual string theory description. The most prominent example of such
a duality relates 4D super Yang-Mills (SYM) theory with ${\cal
N}=4$ supersymmetries to type IIB string theory in the
Anti-deSitter background $AdS_5 \times S_5$. Through the
correspondence, the gauge theoretic BFKL
Pomeron~\cite{Kuraev:1976ge,Kuraev:1977fs,Balitsky:1978ic} gets
related to graviton on the string theory
side~\cite{Kotikov:2004er,Brower:2006ea}.

In~\cite{Bartels:2008zy} and~\cite{Bartels:2009sc} we have
examined this correspondence in some detail. Stimulated by QCD
where $\gamma^*\gamma^*$ scattering provides a safe framework for
investigating the BFKL Pomeron, we have studied the elastic
scattering of two $R$-currents~\cite{CaronHuot:2006te} in ${\cal
N}=4$ SYM theory. On the weak coupling side, the high energy
scattering amplitude factorizes into the current impact factors
and the BFKL Green's function. In~\cite{Bartels:2008zy} the
$R$-current impact factor has been calculated to leading order.
The BFKL Green's function is known also in
NLO~\cite{Fadin:1998py,Ciafaloni:1998gs,Camici:1997ij}. In the
strong coupling region, the method of calculating leading order
correlations function was defined in~\cite{Witten:1998qj}. It
involves the summation of Witten diagrams containing supergravity
fields which live on the $AdS_5$ space. Our calculation of the
high energy behavior of Witten diagrams has shown that the
scattering amplitudes for infinite 't Hooft coupling $\lambda$
also come as a convolution of impact factors and an exchange
propagator, just as in the weakly coupled theory. The convolution
is defined through an integration over the radial direction of the
$AdS_5$ geometry. As a result of our calculation, we have obtained
an expression for the $R$-current impact factor at $\lambda \to
\infty$. Corrections of the order $1/\lambda$ require string
theory calculations. As to the exchanged graviton, Witten diagrams
in the Regge limit yield a power law behavior ${\cal A}_{graviton}
\sim s^j$, with $j=2$ being the spin of the graviton. The higher
order corrections to the graviton trajectory
\be j\ = \ 2-\frac{4+\nu^2}{2\sqrt{\lambda}} + {\cal
O}(\frac{1}{\lambda})\,, \ee
cannot be derived from Witten diagrams, and they have been deduced
from other lines of arguments~\cite{Kotikov:2004er,Brower:2006ea}.
In~\cite{Cornalba:2009ax} a representation for the Regge limit of
four current correlators has been suggested which would allow to
interpolate between weak and strong limits. We have not attempted
to cast our result for the Witten diagram into this form.

Within QCD, it is well known that the BFKL Pomeron violates
unitarity bounds since it grows as ${\cal A}_{BFKL} \sim i s^{j}$
with $j=1+\omega_{BFKL}$ at very high energies. Consequently the
Pomeron must be tamed by suitable corrections. Elaborate
calculations have been performed in order to identify the relevant
corrections within perturbation theory. An example arises in the
context of deep inelastic electron proton scattering at small $x$
(which is related to the elastic scattering of a virtual photon on
the proton). It has been argued that the most important
corrections to the BFKL exchange are given by 'fan' diagrams (an
example is shown in fig.~\ref{fig:triplereggeQCD}a) which contain
the triple Pomeron vertex. This vertex describes the splitting of
one BFKL Pomeron into two Pomerons. A derivation of this result is
obtained by considering, first, the scattering of the virtual
photon on two (weakly coupled) nucleons and, then, closing the two
BFKL Pomerons at the lower end by integrating over the
'diffractive' squared mass $M^2$ (fig.~\ref{fig:triplereggeQCD}b).
As a key feature, the fan diagram in
fig.~\ref{fig:triplereggeQCD}a. contains, in its lower part, the
exchange of two BFKL Pomerons which comes with a minus sign
relative to the single BFKL exchange. At high energies, double
Pomeron exchange grows as ${\cal A}_{double \,\, BFKL} \sim -i
s^{1+2\omega_{BFKL}}$, and thus starts to weaken the growth of the
single BFKL exchange. In preparation for extending this discussion
to ${\cal N}=4$ SYM theory, one may replace the two nucleons at
the bottom by virtual photons. In this way, the essential
amplitude to be studied, becomes the six-point electromagnetic
current correlator, evaluated in the triple Regge limit. It is a
remarkable feature of QCD that the two lower Pomerons do not
couple directly to the upper impact factor. Such a 'direct'
coupling would correspond to the eikonal approximation. The
absence of this direct coupling in the leading logarithmic
approximation of QCD means that the eikonal picture is not
supported.

Turning to ${\cal N}=4$ SYM theory, the analogous correlator is
the six-point correlator of $R$-currents. Our comments on QCD
suggest to investigate, as a first step of addressing the
unitarization, the six-point $R$-current correlator in the limit
$s_1 \sim s_2 \gg M^2$.
\begin{figure}
\begin{center}
{\epsfig{file=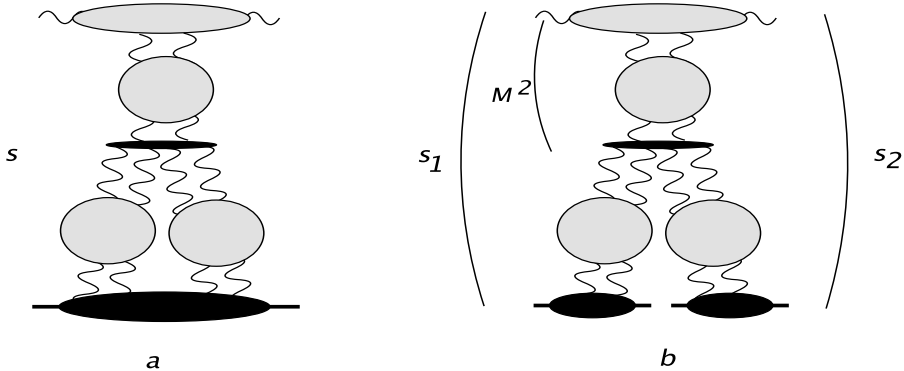,width=12cm,height=5cm}}
\end{center}
\caption{Unitarity corrections in QCD: (a) a fan diagram; (b) the
six-point function} \lab{fig:triplereggeQCD}
\end{figure}
\begin{figure}
\begin{center}
{\epsfig{file=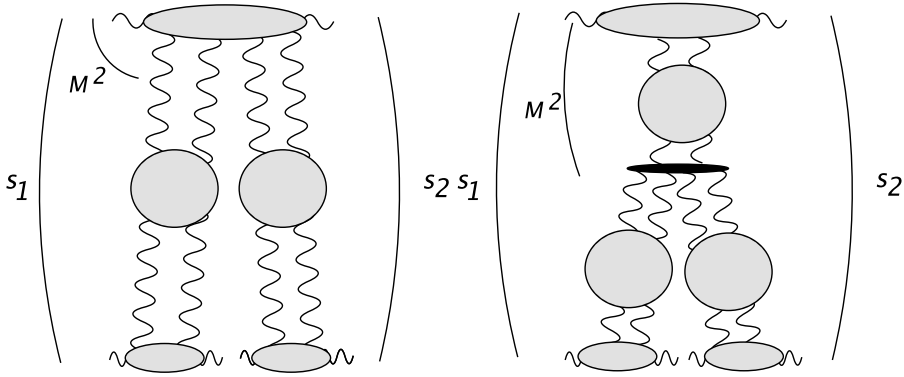,width=12cm,height=5cm}}
\end{center}
\caption{High energy limit of the six-point function in ${\cal N}=4$ SYM}
\lab{fig:triplereggeSYM}
\end{figure}
In the weak coupling limit, this high energy limit of the
six-point $R$-current correlator in ${\cal N}=4$ SYM theory has
been studied in~\cite{Bartels:2009ms}. The main result is
illustrated in fig.~\ref{fig:triplereggeSYM}. At high energies,
the six-point amplitude can be written as a sum of several
pieces~\cite{Brower:1974yv}; each of them corresponds to a
distinct set of simultaneous energy discontinuities, in agreement
with the Steinmann relations. For our discussion we are interested
only in those terms which contribute to the discontinuity in the
energies $s_1$, $s_2$ and in the square of the diffractive mass,
$M^2$. In the leading log approximation, the triple Pomeron vertex
(fig.~\ref{fig:triplereggeSYM}, right figure) is the same as in
QCD. The amplitude corresponding to this diagram has the form
\be {\cal A}_{3\to3}^{triple}\ = \ \frac{s_1s_2}{M^2}\int\int\int
\frac{d\omega d\omega_1 d\omega_2}{(2\pi i)^3}
s_1^{\omega_1} s_2^{\omega_2} (M^2)^{\omega-\omega_1 -\omega_2}\nonumber\\[2mm]
\xi(\omega_1) \xi(\omega_2) \xi(\omega,\omega_1,\omega_2)
F(\omega,\omega_1,\omega_2)\,, \ee
where the signature factors are given by
\be \xi(\omega) \ = \ -\pi \frac{e^{-i pi \omega}-1}{\sin \pi
\omega}\quad ,\quad \,\,\, \xi(\omega,\omega_1,\omega_2) \ = \
-\pi \frac{e^{-i \pi(\omega-\omega_1 -\omega_2)} -1}{\sin
\pi(\omega-\omega_1 -\omega_2)}\,, \ee
and
\be F(\omega,\omega_1,\omega_2)\ = \ \Phi(Q^2)\otimes
G(\omega)\otimes V \otimes G(\omega_1 )\otimes \Phi(Q_A^2) \otimes
G(\omega_2)\otimes\Phi(Q_B^2)\,. \ee
Here $\otimes$ denotes the integration over transverse momenta,
$G(\omega)$ is the BFKL Green's function, $\Phi$ is the impact
factor presented in ~\cite{Bartels:2008zy}, and details on the
triple Pomeron vertex $V$ can be can be found
in~\cite{Bartels:2009ms}. The discontinuity of this six-point
function across the cut in $M^2$ leads to the cross section of the
diffractive scattering process (in the notations of QCD)
$\gamma^*+\gamma^* \to M_X +\gamma^*$. Since $M^2$ is large, we
obtain a contribution of 'large diffractive masses'. In all three
$\omega$ variables, the leading singularity is given by the BFKL
characteristic functions
\ba \omega\ = \ \omega_1\ = \ \omega_2\ = \  \alpha_s
\chi(\nu=0,n=0)\ = \ \frac{4N_c \alpha_s \ln 2}{\pi}\ . \ea
As an important feature of ${\cal N}=4$ SYM theory we find an
extra contribution (see fig.~\ref{fig:triplereggeSYM}, left
figure) where the two BFKL exchanges couple directly to the upper
$R$-currents. The presence of this 'direct' coupling, which is
absent in QCD and might be viewed as a support of eikonalization
in ${\cal N}=4$ SYM theory, can be traced back to the fact that
fermions (and scalars) belong to the adjoint representation. The
corresponding scattering amplitude is of the form
\ba {\cal A}_{3\to3}^{direct}\ = \ s_1s_2 \int \int
\frac{d\omega_1 d\omega_2}{(2\pi i)^2} s_1^{\omega_1}
s_2^{\omega_2}\xi(\omega_1) \xi(\omega_2)
F(M^2;\omega_1,\omega_2)\,, \ea
where
\be F(M^2;\omega_1,\omega_2) \ = \  \Phi(Q^2;M^2)^{direct}
\otimes G(\omega_1) \otimes \Phi(Q_A^2) \otimes G(\omega_2)
\otimes \Phi(Q_B^2)\,. \lab{eq:Fdirect} \ee
An expression for the new impact factor $\Phi(M^2; Q^2))^{direct}$
which describes the coupling of the two BFKL Pomerons to the upper
$R$-current can be found in~\cite{Bartels:1994jj,Bartels:1996ne}
and~\cite{Bartels:2009ms,BEHM} \footnote{Ref.~\cite{BEHM}
discusses the $M^2$ discontinuity of the two-Pomeron impact
factor. In order to obtain the full impact factor from this
discontinuity, one writes a (unsubtracted) dispersion relation.}.
For large $M^2$, this impact factor falls off as $M^{-4}$. For the
diffractive cross section one takes the $M^2$-discontinuity of the
six-point amplitude, i.e. the $M^2$-discontinuity of the impact
factor $\Phi(M^2; Q^2))^{direct}$. The latter falls off as
$M^{-8}$, i.e.\ it contributes to the region of small diffractive
masses.

In the present paper we continue the investigtion of the high
energy limit in the strongly coupled theory using Witten diagrams.
Our main interest now is in the six-point $R$-current correlators.
In the triple Regge limit, the amplitude is dominated by
$t-$channel exchanges of gravitons. The relevant diagrams are
shown in figs.~\ref{fig:gravboson} and \ref{fig:grav3}. There is
an obvious correspondence between the two contributions on the
weak (fig.~\ref{fig:triplereggeSYM}) and on the strong coupling
side (figs.~\ref{fig:gravboson} and \ref{fig:grav3}, left diagram).
\begin{figure}
\begin{center}
{\epsfysize5cm \epsfbox{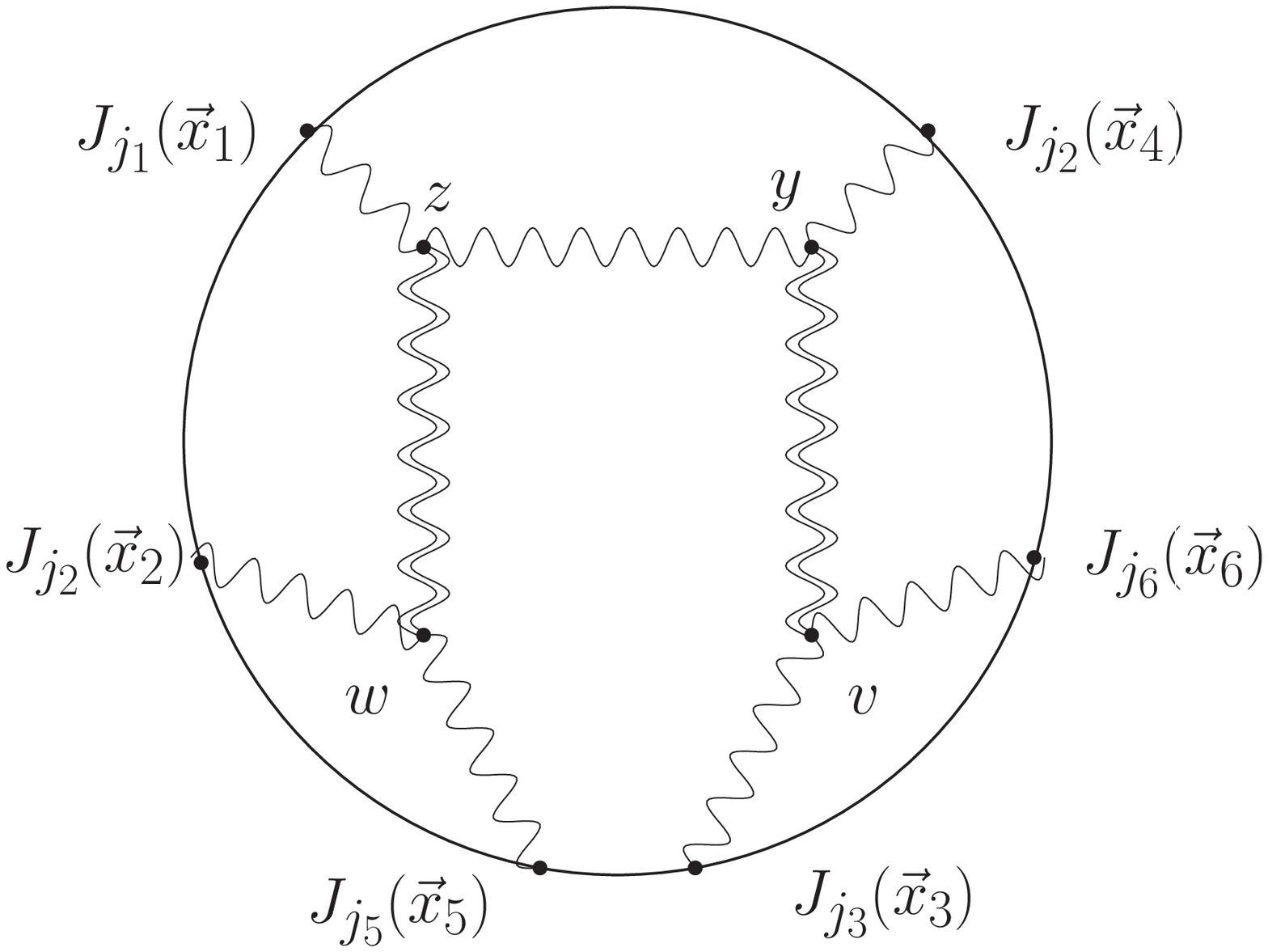}\qquad\epsfbox{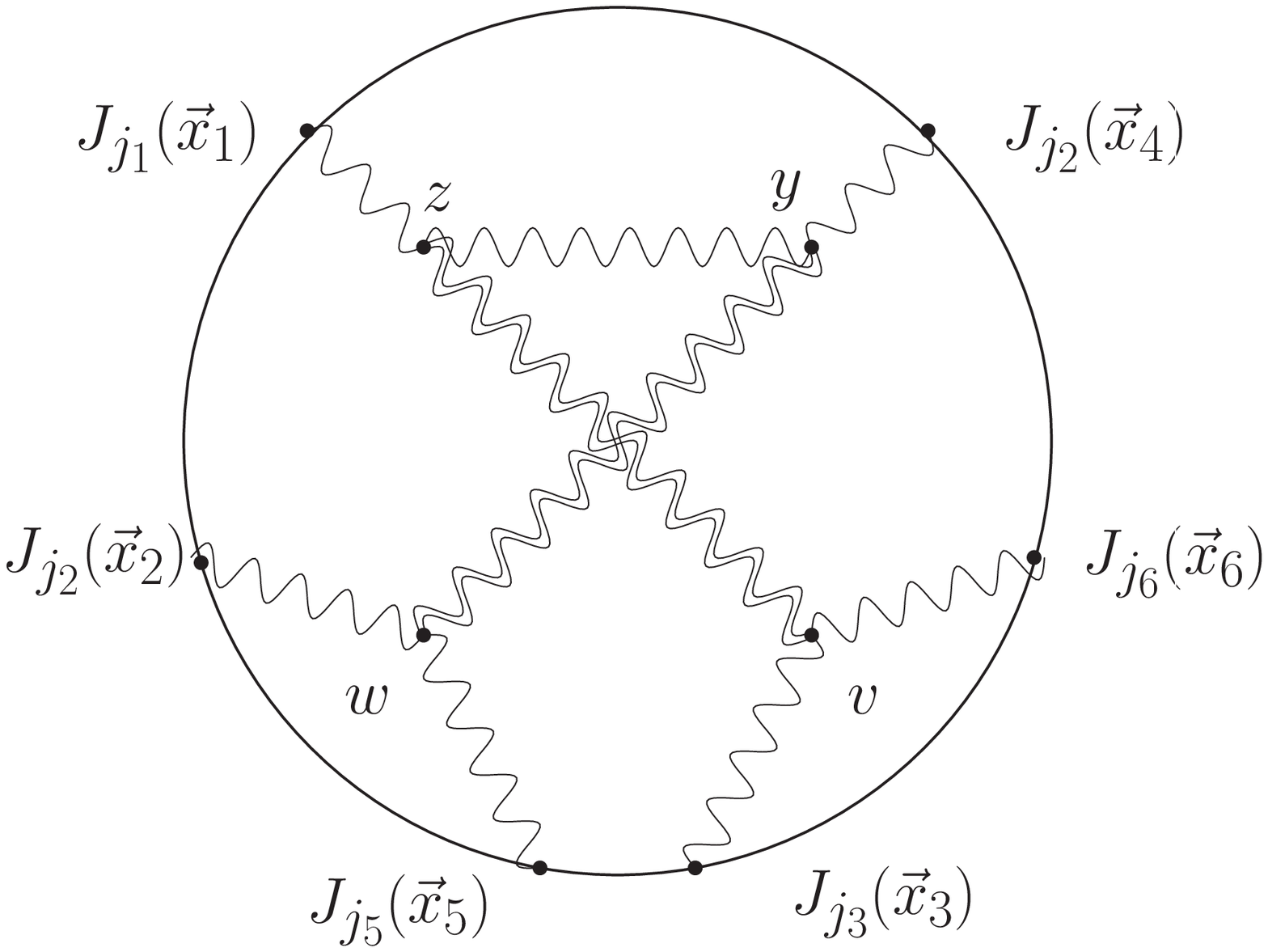}}
\end{center}
\caption{Witten diagrams for the two graviton exchange
in the $t-$channel}
\lab{fig:gravboson}
\end{figure}
These Witten diagrams will be considered as the strong coupling
analogue of our weak coupling results obtained in ${\cal N}=4$ SYM
theory .

Our article is organized as follows. Section \ref{sc:6pj} is
devoted to a brief review of our notation used
in~\cite{Bartels:2009sc}. In section \ref{sc:T2G} we present
computations of the scattering amplitude with the two $t-$channel
gravitons and one intermediate $R$-boson carrying mass $M$
(fig.~\ref{fig:gravboson}). We rewrite the amplitude to momentum
space and perform the high energy limit. The amplitude is found to
be proportional to the square of two large energy variables,
namely $s_1^2 s_2^2$. The planar graph (left part of
fig.~\ref{fig:gravboson}) has a cut for positive $M^2$, starting
at $M^2=0$, and, for large $M^2$ (triple Regge limit), falls off
as $M^{-2}$. Correspondingly, the crossed graph (right part of
fig.~\ref{fig:gravboson}) has a cut for negative values of $M^2$.
Finally, in section \ref{sc:TGV} we consider the correlation
function with the triple graviton vertex (fig.~\ref{fig:grav3}).
In the triple Regge limit, the expected contribution to the triple
Regge behavior $\sim (s_1/M^2)^{j_1} (s_2/M^2)^{j_2} (M^2)^{j}$
with $j=j_1=j_2=2$ vanishes. Instead, we find contributions
proportional to $s_1^2$, $s_2^2$, and $s_1 s_2$.
\begin{figure}
\begin{center}
{\epsfysize5cm \epsfbox{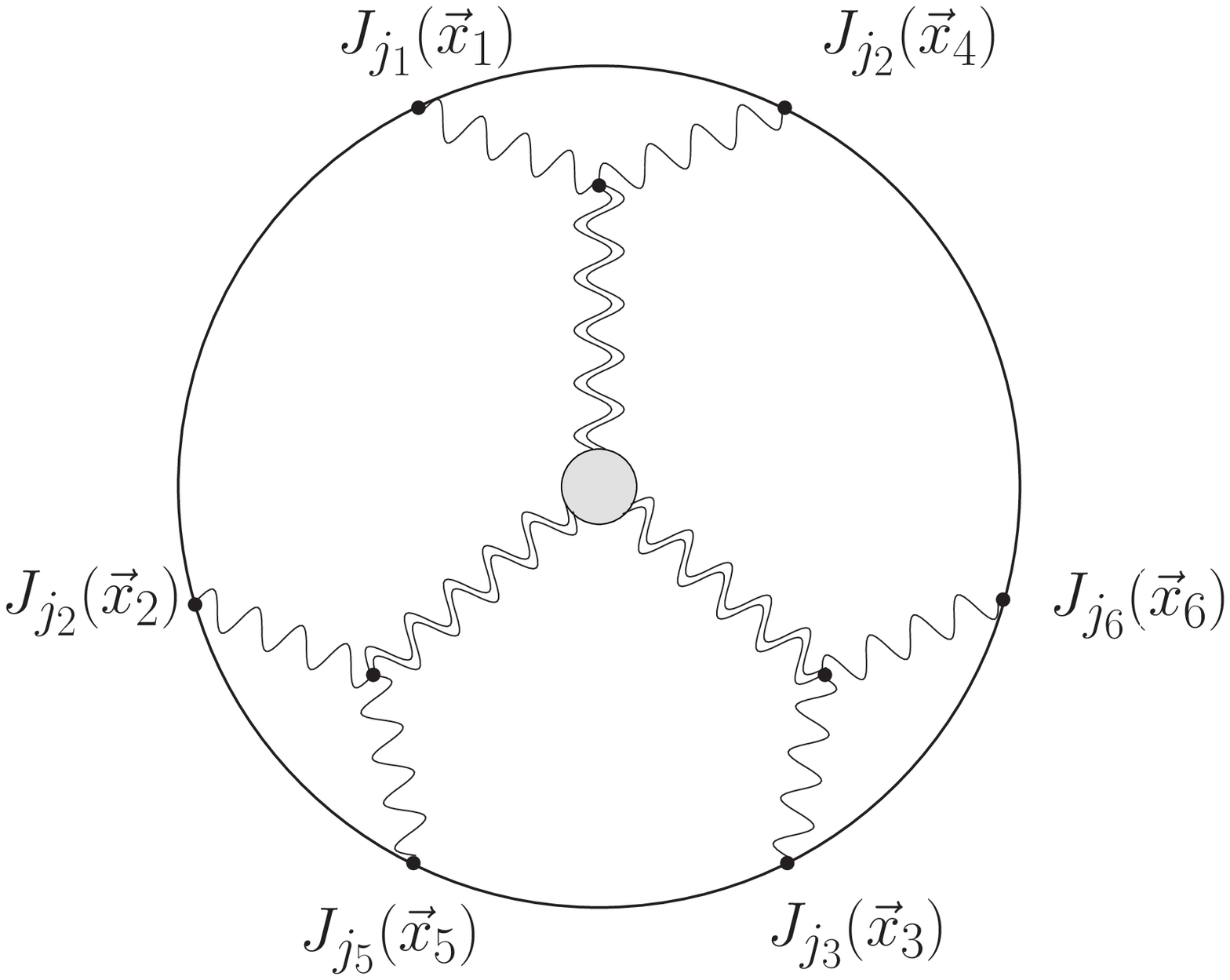}\qquad\epsfbox{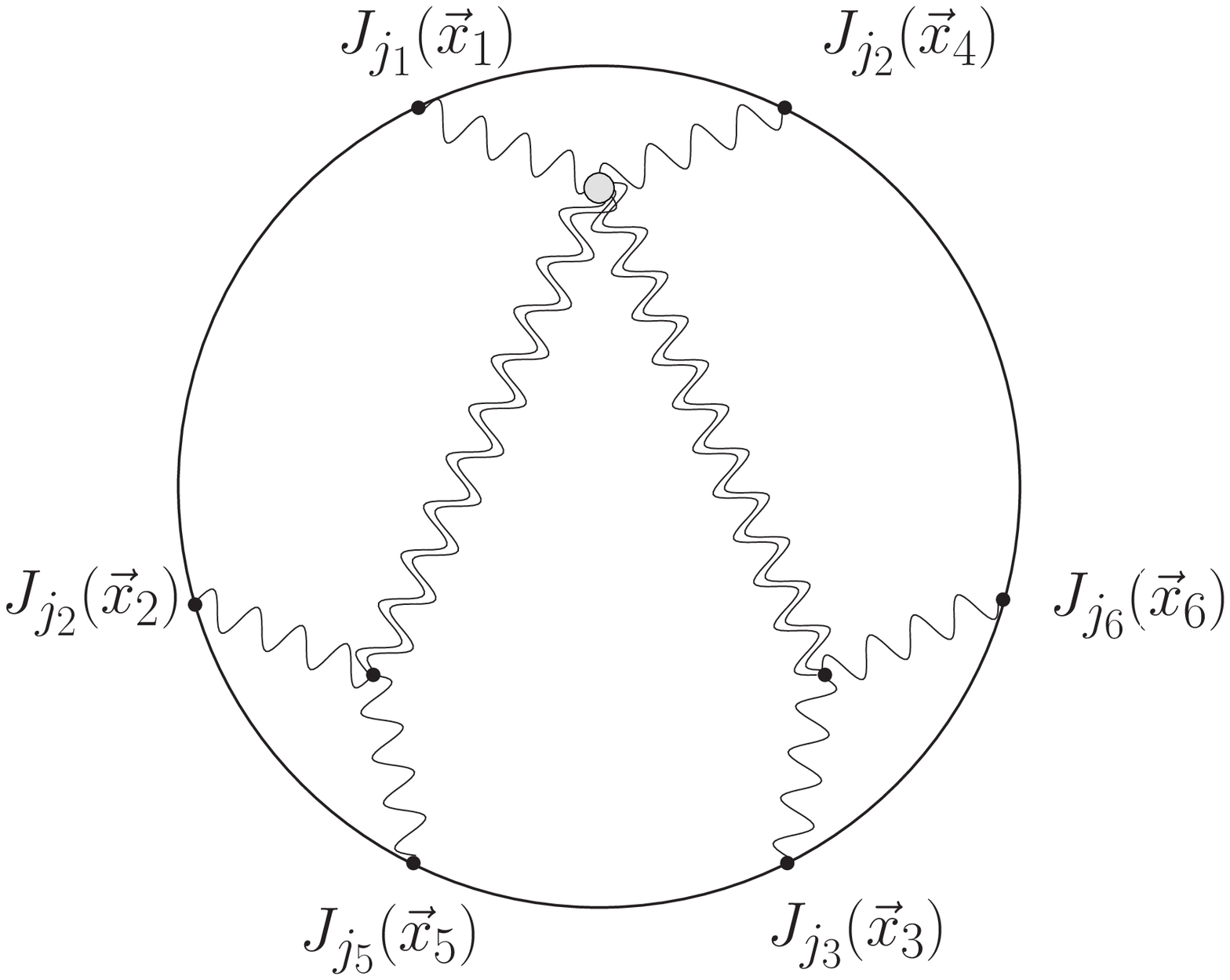}}
\end{center}
\caption{Left figure: triple graviton exchange with the triple graviton 
vertex. Right figure:  two graviton exchange with the direct coupling of 
two gravitons and two bosons.}
\lab{fig:grav3}
\end{figure}

\section{Six-point correlation functions at strong coupling}
\lab{sc:6pj}

Let us consider ${\cal N}=4$ super Yang-Mills (SYM) theory in
 four dimensional Euclidean space.
The Fourier
transform of the six-point correlator reads as
\be
 (2 \pi)^4 \delta(\sum_{i=1}^{6}\vec p_i)
A_{j_1 j_2 j_3 j_4 j_5 j_6}(\vec p_i) \ =\ \int \left(
\prod_{i=1}^6 d^4 x_i \e^{-i \vec p_i \cdot \vec x_i} \right)
 \langle \prod_{a=1}^6 J_{j_a}(\vec x_a) \rangle\, . \lab{eq:amp} \ee
By $J_{j}$ we denote $R$-currents
 with $j$ labelling the spacial directions,
$j=1,\dots,d=4$. The $\vec{x}=(x_1,x_2,x_3,x_4)$ stands for the four
dimensional Euclidean vector (the value $j=0$ refers to the fifth coordinate).

We use the same notations as in ~\cite{Bartels:2009sc}. Starting
with the Euclidean notation $\vec p =(p_1,p_2,p_3,p_4)$ and $|\vec
p| = \sqrt{\vec{p}^2}$, the Wick rotation continues $|\vec p|^2=
\vec p^2 \to - p^2=- p_4^2 +p_1^2 +p_2^2 +p_3^2$ in Minkowski
space. In the high energy limit, our scattering amplitude depends
upon the energies $s_1$, $s_2$, the diffractive mass squared
$M^2$, and the momentum transfers $t_1$, $t_2$, and $t$.
Furthermore, $|\vec p_1|$, $|\vec p_2|$, $|\vec p_3|$, ($|\vec
p_4|$ ,$|\vec p_5|$ and $|\vec p_6|$) are the virtualities of the
incoming (outgoing) currents. In Euclidean notation we have
\ba &&s_1\ =\ -(\vec p_1+\vec p_2)^2\,, \quad s_2\ =\ -(\vec
p_1+\vec p_3)^2\,, \quad M^2 \ = \  -(\vec p_1+\vec p_2+\vec
p_5)^2\,, \quad
\nonumber \\[2mm]
&&-t\ =\ (\vec p_1+\vec p_4)^2\,, \quad
-t_1\ =\ (\vec p_2+\vec p_5)^2\,, \quad
-t_2\ =\ (\vec p_3+\vec p_6)^2\,. \quad
\ea
After Wick rotations, the energy variables $s_1$, $s_2$, and $M^2$
are positive, whereas the momentum transfer variables $t$, $t_1$,
$t_2$ remain negative; the masses of the external currents are
kept negative (space-like), $-|\vec p_i|$. After Wick rotation, we
still continue to use the vector symbol $\vec p$ for the Minkowski
vector $(p_1,p_2,p_3,p_4)$, but now $\vec p ^2 = -p^2$.

The high energy limit is defined as
\ba s_1,\,s_2\,\gg M^2\,-t,\,-t_1,\,-t_2, -|\vec p_i|^2. \ea
For the two graviton exchange diagrams we will keep $M^2$ finite,
whereas for the triple graviton diagram we take the triple Regge
limit where also $M^2$ becomes large.

Finally, we find it convenient to present the scattering amplitude in the helicity
basis. To this end we contract the correlator $A$ with appropriate
polarization vectors
\ba & & {\cal
A}_{\lambda_1\lambda_2\lambda_3;\lambda_4\lambda_5\lambda_6}(|\vec
p_i|;s_1,s_2,M^2;t,t_1,t_2)\ =  \nonumber \\[2mm] &&
\hspace*{2mm} = \sum_{j_i}
 \epsilon_{j_1}^{(\lambda_1)}(\vec p_1)
 \epsilon_{j_2}^{(\lambda_2)}(\vec p_2)
 \epsilon_{j_3}^{(\lambda_3)}(\vec p_3)
 \epsilon_{j_4}^{(\lambda_4)}(\vec p_4)^{\ast}
 \epsilon_{j_5}^{(\lambda_5)}(\vec p_5)^{\ast}
 \epsilon_{j_6}^{(\lambda_6)}(\vec p_6)^{\ast}
{A}_{j_1 j_2 j_3 j_4 j_5 j_6}(\vec p_i) \,,
\ea
where $\lambda_i=L,\pm$ runs through the possible helicities and
we introduced the polarization vectors
$\epsilon_{j}^{\lambda_i}(\vec p_i)$ such that $p_i^{j}
\epsilon_{j}^{(\lambda_i)}(\vec{p_i}) =0$.

In order to calculate the amplitude (\ref{eq:amp}) in the limit of
infinite 't Hooft coupling~\cite{Bartels:2008zy} we make use of
the conjectured AdS/CFT correspondence~\cite{Maldacena:1997re}
between IIB string theory on $AdS_{5}$ space and ${\cal N}=4$
$SU(N_c)$ super Yang-Mills theory. An efficient calculation can
only be performed in the limit of large $N_c$. Moreover the full
string theory on $AdS_{5}$ is well approximated by classical
supergravity when 't Hooft coupling $\lambda=g^2_{YM} N_c$ goes to
infinity.

According to the AdS/CFT correspondence, correlation functions are
related with a classical supergravity action $S_{AdS}$ by
~\cite{Witten:1998qj,Gubser:1998bc}
\be \langle J(1) J(2) \ldots J(n) \rangle_{CFT} \ = \ \omega_n
\frac{\delta^n}{\delta \phi_0(1) \ldots \delta \phi_0(n)}
\exp(-S_{AdS}[\phi[\phi_0]]) \big|_{\phi_0=0} \,,
\lab{eq:AdSCFTcorr} \ee
where the factor $\omega_n$ comes from the relative
normalization~\cite{Chalmers:1998xr} while the sources $\phi_0$ of
operators in super Yang-Mills theory correspond to the boundary
values of supergravity fields in $AdS_{5}$ in the $4$-dimensional
quantum field theory, i.e.\ $\phi|_{\partial AdS}\sim\phi_0$. We
are using the following conventions concerning the Anti-deSitter
space $AdS_{d+1}$. Its Euclidean continuation is parameterized by
$z_0>0$ and $\vec x$ with coordinates $x_i$ enumerated by the
Latin indices $i=1,\ldots,d$. We use the metric
\be ds^2\ = \  \frac1{z_0^2}(d z_0^2+ d \vec x^2)\,, \lab{eq:ds2}
\ee
where $d \vec x^2$ can be related to the metric of Minkowski space
by Wick rotation. The limit $z_0 \to 0$ corresponds to the
boundary of the Anti-deSitter space. The most interesting case is
for $d=4$ which can be related to QCD.

To simplify notation we truncate the $SU(4)$ $R$-current group to
$U(1)_R$. However, our considerations may easily be generalized to
the non-Abelian case. The supergravity action is defined by
\be S\ = \ \frac{1}{2 \kappa_{d+1}^2} \int d^{d+1} z
\sqrt{g}(-{\cal R}+\Lambda)+S_m\,, \lab{eq:Sg} \ee
where ${\cal R}$ is the scalar curvature while the covariant
matter action reads
as~\cite{Freedman:1998tz,Chalmers:1998xr,Cvetic:1999xp,Arutyunov:2000py}
\be S_m\ = \  \frac{1}{2\kappa_{d+1}^2} \int d^{d+1} z \sqrt{g}
\left[\frac{1}{4} F_{\mu \nu} F^{\mu \nu}
-A_\mu J^\mu +\ldots \right]\ .
\lab{eq:Sm} \ee
Here $\kappa_5^2$ is fixed by matching two- and three-point
protected operators~\cite{Freedman:1998tz,Chalmers:1998xr},
while $F_{\mu \nu}$ is the field strength of the gauge field $A$.
Throughout this note, Greek indices refer to the
$(d+1)$-dimensional space, i.e.\ they take values from $0$ to $d$.
Latin subscripts, on the other hand, parameterize directions along
the Euclidean $d$-dimensional boundary of $AdS_{d+1}$.
Contractions of the full metric (\ref{eq:ds2}) are denoted with
upper and lower indices while contractions of both lower indices
denotes simple summation with Kronecker delta.

After these technical preparations we can now begin to evaluate
the high energy limit of our six-point correlator at strong
coupling, where supergravity on AdS is believed to provide an
accurate description. To this end we make use of a very convenient
and intuitive diagrammatic procedure that was first proposed by
Witten~\cite{Witten:1998qj} and then developed further by many
other authors. It relies on summing diagrams which in our case
contain only three basic building blocks, namely the bulk-to-bulk
propagators for the graviton and the gauge $R$-bosons as well as
the bulk-to-boundary $R$-boson propagator. They are connected by
vertices defined in eqs.\ (\ref{eq:Sg}) and (\ref{eq:Sm}). In the
high energy limit it is enough to analyze diagrams plotted in
figs.~\ref{fig:gravboson}, \ref{fig:grav3}.

\section{Two Graviton exchange: Low diffractive masses}
\lab{sc:T2G}

In this section we analyze two Witten diagrams depicted in
fig.~\ref{fig:gravboson}. These will later turn out to contain all
leading order contributions to the high energy limit of the full
amplitude. After a very detailed discussion of the first diagram
we can obtain the contribution from the second diagram through
analytic continuation. The results are spelled out in eqs.\
(\ref{eq:XPPP}) and (\ref{eq:XPPPcr}). They involve a new impact
factor, defined in eq.\ (\ref{eq:2gravimpact}), whose properties
shall be analyzed in subsection 3.3. The final subsection is then
devoted to a study of the deep inelastic limit of the amplitude.

\subsection{The Momentum space representation}

We start from the expression for the two graviton exchange
in configuration space. Its contribution to the six current matrix element
is\footnote{The
correlation functions and amplitudes are calculated up to multiplicative constants, which can be easily restored from the action (\ref{eq:Sg}).}
\ba I^{\rm 2G,planar}&=& \int \frac{d^{d+1}y}{y_0^{d-3}} \int
\frac{d^{d+1}v}{v_0^{d-3}} \int \frac{d^{d+1}w}{w_0^{d-3}} \int
\frac{d^{d+1}z}{z_0^{d-3}} \tilde T_{(14) \mu \nu,\rho
\sigma}(z,y)
 \\[2mm] &&
G_{\mu \nu; \mu' \nu'}(z,w) G_{\rho \sigma; \rho' \sigma'}(y,v)
\tilde T_{(25) \mu' \nu'}(w) \tilde T_{(36) \rho' \sigma'}(v)\,,
\nonumber \ea where the stress-energy tensor \ba \tilde T_{(14)\mu
\nu}&=& z_0^2
\partial_{[\mu} G_{\lambda] \rho_1}(z,\vec x_1)
\partial_{[\nu} G_{\lambda] j_4}(z,\vec x_4)
+
z_0^2
\partial_{[\nu} G_{\lambda] \rho_1}(z,\vec x_1)
\partial_{[\mu} G_{\lambda] j_4}(z,\vec x_4)
\nonumber \\[2mm]
&& \hspace*{2cm} - \frac1{2} z_0^2 \delta_{\mu \nu}
\partial_{[\alpha} G_{\beta] \rho_1}(z,\vec x_1)
\partial_{[\alpha} G_{\beta] j_4}(z,\vec x_4)
\,.
\ea
In the high energy limit, the highest contribution comes from the
first two terms. For the coupling of the two gravitons to the
upper currents one can define the double stress-energy tensor
\ba \tilde T_{(14)\mu \nu, \rho \sigma} &=& (\delta_{\mu
\mu'}\delta_{\nu \nu'} +\delta_{\mu \nu'}\delta_{\nu \mu'} )
(\delta_{\rho \rho'}\delta_{\sigma \sigma'} +\delta_{\rho
\sigma'}\delta_{\sigma \rho'} )
\nonumber \\[2mm]
&&
[
z_0^2
y_0^2
\partial_{z_{[\mu'}} G_{\lambda] j_1}(z,\vec x_1)
\partial_{y_{[\rho'}}\partial_{z_{[\nu'}} G_{\lambda] \tau]}(z,y)
\partial_{y_{[\sigma'}} G_{\tau],j_4}(y,\vec x_4)
\nonumber \\[1mm]
&&-
\frac1{2}
z_0^2
y_0^2
\delta_{\mu' \nu'}
\partial_{z_{[\alpha}} G_{\lambda] j_1}(z,\vec x_1)
\partial_{y_{[\rho'}}\partial_{z_{[\alpha}} G_{\lambda] \tau]}(z,y) G
\partial_{y_{[\sigma'}} G_{\tau],j_4}(y,\vec x_4)
\nonumber \\[1mm]
&&
-
\frac1{2}
y_0^2
z_0^2
\delta_{\sigma' \rho'}
\partial_{z_{[\nu'}} G_{\lambda] j_1}(z,\vec x_1)
\partial_{y_{[\beta}}\partial_{z_{[\mu'}} G_{\lambda] \tau]}(z,y)
\partial_{y_{[\beta}} G_{\tau] j_4}(y,\vec x_4)
\nonumber \\[1mm]
&&+
\frac1{4}
y_0^2
z_0^2
\delta_{\sigma' \rho'}
\delta_{\mu' \nu'}
\partial_{z_{[\alpha}} G_{\lambda] j_1}(z,\vec x_1)
\partial_{y_{[\beta}} \partial_{z_{[\alpha}} G_{\lambda] \tau]}(z,y)
\partial_{y_{[\beta}} G_{\tau] j_4}(y,\vec x_4) ]\,.
\ea
In the high energy limit, only the first term contributes to the
leading power in energy. The expressions for the propagators are
listed in Ref.~\cite{Bartels:2009sc}.

Let us now specify $d=4$. Using the expressions for the
propagators presented in Ref.~\cite{Bartels:2009sc} we rewrite the
formulae in the momentum space. We define Fourier transform of
stress-energy tensors as
\ba \tilde T_{(14)\mu\nu}(z) &=&
\frac1{(2 \pi)^{8}} \int d^4 p_1\int d^4 p_4 \e^{-i \vec p_1 \cdot
(\vec z-\vec x_1)} \e^{-i \vec p_4 \cdot (\vec z-\vec x_4)}
T_{(14)\mu\nu}(z_0;\vec p_1, \vec p_4)\,, \ea
and
\ba \tilde T_{(14)\mu\nu,\rho \sigma}(z,y) &=& \frac1{(2
\pi)^{12}} \int d^4 p_1\int d^4 p_4 \int d^4 p\,  \e^{-i \vec p_1
\cdot (\vec z-\vec x_1)} \e^{-i \vec p_4 \cdot (\vec y-\vec x_4)}
\e^{- i \vec p
\cdot (\vec y-\vec z)} \nonumber \\[2mm] && \hspace*{5cm} T_{(14)\mu\nu,\rho
\sigma}(z_0,y_0;\vec p_1,\vec p, \vec p_4) \,. \ea
This gives
\ba T_{(14)\mu \nu}(z_0;\vec p_1, \vec p_4)&\approx& - z_0^2
{p_1}_{[\mu} G_{\lambda] \rho_1}(z_0,\vec p_1) {p_4}_{[\nu}
G_{\lambda] j_4}(z_0,\vec p_4) \nonumber \\[2mm] && - z_0^2
{p_1}_{[\nu} G_{\lambda] \rho_1}(z_0,\vec p_1) {p_4}_{[\mu}
G_{\lambda] j_4}(z_0,\vec p_4) \,, \ea
with ${p_k}_0 \equiv i {\partial}_{z_0}$ and
\ba T_{(14)\mu \nu, \rho \sigma}(z_0,y_0;\vec p_1,\vec p, \vec
p_4)&\approx& z_0^2 y_0^2 {p_1}_{{[\mu}} G_{\lambda] j_1}(z_0,\vec
p_1) {\bar p}_{{[\rho}}{ p}_{{[\nu}} G_{\lambda]
\tau]}(z_0,y_0;\vec p) {p_4}_{{[\sigma}} G_{\tau],j_4}(y_0,\vec
p_4)
\\[2mm]
&&+
z_0^2
y_0^2
{p_1}_{{[\nu}} G_{\lambda] j_1}(z_0,\vec p_1)
{\bar p}_{{[\rho}}{ p}_{{[\mu}} G_{\lambda] \tau]}(z_0,y_0;\vec p)
{p_4}_{{[\sigma}} G_{\tau],j_4}(y_0,\vec p_4)
\nonumber \\[2mm]
&&+
z_0^2
y_0^2
{p_1}_{{[\mu}} G_{\lambda] j_1}(z_0,\vec p_1)
{\bar p}_{{[\sigma}}{ p}_{{[\nu}} G_{\lambda] \tau]}(z_0,y_0;\vec p)
{p_4}_{{[\rho}} G_{\tau],j_4}(y_0,\vec p_4)
\nonumber \\[2mm]
&&+
z_0^2
y_0^2
{p_1}_{{[\nu}} G_{\lambda] j_1}(z_0,\vec p_1)
{\bar p}_{{[\sigma}}{ p}_{{[\mu}} G_{\lambda] \tau]}(z_0,y_0;\vec p)
{p_4}_{{[\rho}} G_{\tau],j_4}(y_0,\vec p_4) \,,
\nonumber
\ea
with $\partial_{\vec z_i} = -i p_i$, $\partial_{\vec y_i} = -i
\bar p_i$, $p_0=i\partial_{z_0}$, $\bar p_0=i\partial_{y_0}$,
${p_1}_0 \equiv i {\partial}_{z_0}$, ${p_4}_0 \equiv i
{\partial}_{y_0}$. In the above formulae the {\em approximation}
indicates that we omit terms which, in the high energy limit, are
power suppressed.

Finally, our expression in the four-dimensional momentum space
takes the following form
\ba  & & (2 \pi)^{4} \delta^{(4)}(\sum_i \vec p_i) {A}^{\rm
2G,planar}_{j_1 j_2 j_3 j_4 j_5 j_6}(\vec p_i) \ = \
\left(\prod_{j=1}^{6} \int d^4 x_j \e^{-i \vec x_j \cdot \vec
p_j}\right)
 I^{\rm 2G,planar}\ =
\nonumber \\[2mm]
&=&  {(2 \pi)^{4}} \delta^{(4)}(\sum_i \vec p_i) \int_0^{\infty}
\frac{d y_0}{y_0} \int_0^{\infty} \frac{d v_0}{v_0}
\int_0^{\infty} \frac{d w_0}{w_0} \int_0^{\infty} \frac{d
z_0}{z_0} \, T_{(14)\mu\nu,\rho \sigma}(z_0,y_0;\vec p_1, \vec
p_1+ \vec p_2+\vec p_5, \vec p_4)
\nonumber \\[2mm]&& \hspace*{3mm}
G_{\mu \nu; \mu' \nu'}(z_0,w_0; \vec p_2+\vec p_5)
T_{(25)\mu'\nu'}(w_0;\vec p_2, \vec p_5)
G_{\rho \sigma; \rho'
\sigma'}(y_0,v_0;\vec p_3+\vec p_6)\nonumber T_{(36)\rho'
\sigma'}(v_0;\vec p_3, \vec p_6)\,. \ea

\subsection{The high energy limit}
In the high energy limit, the leading contribution can be obtained
exactly in the same way as it was done for four point
functions~\cite{Bartels:2009sc}. For the incoming $R$-boson
propagators, the only important parts are those proportional to
$p_k$, namely
\ba {p}_{[k} G_{l] j}(z_0, \vec p) &=&
 z_0
(
{p}_{k} \delta_{l j }
-
{p}_{l} \delta_{k j }
)
 |\vec p| K_{1}(z_0|\vec p|)
\approx
 z_0
{p}_{k} \delta_{l j }
 |\vec p| K_{1}(z_0|\vec p|)\,,
\lab{eq:pG1}
\ea
\ba
{p}_{[k} G_{0] j}(z_0, \vec p) &=&
i z_0
(
\delta_{k j }
 |\vec p|^2
- {p}_{j}
 {p}_{k}
)
 K_{0}(z_0|\vec p|)
\approx
-i z_0
 {p}_{j}
 {p}_{k}
 K_{0}(z_0|\vec p|)\,,
\lab{eq:pG0} \ea
where $\partial_{\vec z_i} = -i p_i$, $p_0=i\partial_{z_0}$.
Making use of the Ward identity, i.e.\ shifting the polarization
vectors (listed in~\cite{Bartels:2009sc}, Appendix A), we can
remove terms without $p_k$. To simplify the notation of the
bulk-to-bulk $R$-boson propagator we introduce
\ba {\cal K}_a (z_0,y_0;|\vec p|) &=& \sum_{k=0}^{\infty}
\frac{2^{-2k-1}}{\Gamma(k+2)\Gamma(k+1)} \left(\frac{z_0 y_0 |\vec
p|}{\sqrt{z_0^2+y_0^2}}\right)^{2k+a} K_{2k+a}(|\vec p|
\sqrt{z_0^2+y_0^2})\,, \ea and \be \tilde {\cal K}_a(z_0,y_0;|\vec
p|) \ = \ \sum_{k=0}^{\infty}
\frac{2^{-2k-a}}{\Gamma(k+1+a)\Gamma(k+1)} \left(\frac{z_0 y_0
|\vec p|}{\sqrt{z_0^2+y_0^2}}\right)^{2k+a} K_{2k+a}(|\vec p|
\sqrt{z_0^2+y_0^2})\,. \lab{eq:tKa} \ee This allows us to rewrite
the bulk-to-bulk $R$-boson propagators as \ba G_{\mu j}(z_0, y_0,
\vec p)&=& \ft1{8}\delta_{\mu j} z_0 y_0 \tilde {\cal K}_1
-\ft{i}{8} p_j \delta_{\mu 0} z_0 y_0^2 {\cal K}_0\,, \ea and \ba
G_{\mu 0}(z_0, y_0, \vec p)&=& \ft1{8}\delta_{\mu 0} (z_0^2+
y_0^2) \tilde {\cal K}_1 - \ft1{8}\delta_{\mu 0} z_0 y_0 \tilde
{\cal K}_0 +\ft{i}{8} p_j \delta_{\mu j} z_0^2 y_0 {\cal K}_0\,.
\ea
Furthermore, in the high energy limit
 the leading term of the graviton propagator is given by
\ba G_{\mu \nu; \mu' \nu'}(z_0,w_0,\vec p) &\approx& (\delta_{\mu
\mu'}\delta_{\nu \nu'}+\delta_{\mu \nu'}\delta_{\nu \mu'} ) {\cal
G}(z_0,w_0;\vec p)\,, \ea with \ba {\cal G}(z_0,w_0;\vec
p)&\equiv& \tilde {\cal K}_{a=2}(z_0,w_0;|\vec p|)\,. \ea
To calculate the scattering amplitude we have to contract the
resulting expression with the polarization vectors, namely
\be {\cal A}^{\rm 2G,planar}_{\lambda_1 \lambda_2 \lambda_3;
\lambda_4 \lambda_5 \lambda_6 } \ = \ {\sum}_{j_i} \prod_{a=1}^3
\epsilon_{j_a}^{(\lambda_a)}(\vec p_a) \ ({A}^{\rm
2G,planar})_{j_1 j_2 j_3; j_4 j_5 j_6} \ \prod_{b=4}^6
 \epsilon_{j_b}^{(\lambda_b)}(\vec p_b)^{\ast}
\,. \ee
Substituting the expressions for the propagators, the double
stress-energy tensor reads as
\ba T_{(14)\mu \nu, \rho \sigma}(z_0,y_0;\vec p_1,\vec p;\vec p_4)
&\approx& - \ft{1}{8} z_0^4 y_0^4 {p_1}_{k_1}
 p_{k_2}
p_{k_3}
 {p_4}_{k_4}
(\delta_{\mu k_1}\delta_{\nu k_3} +\delta_{\mu k_3}\delta_{\nu
k_1} ) (\delta_{\rho k_2}\delta_{\sigma k_4} +\delta_{\rho
k_4}\delta_{\sigma k_2}) \nonumber \\[2mm] && \sum_{m=0,1} W^{m}_{j_1
j_4}(\vec p_1,\vec p_4) K_{m}(z_0|\vec p_1|) \tilde {\cal
K}_{m}(z_0,y_0;|\vec p|) K_{m}(y_0|\vec p_4|)\,, \lab{eq:dbT} \ea
where we have introduced the vector \be \vec p \ = \ \vec p_1+
\vec p_2+\vec p_5. \lab{eq:vectorp} \ee The tensor part, namely
\be W^{m}_{j_1 j_4}(\vec p_1,\vec p_4) \ = \ (\delta_{j_1 j_4}
|\vec p_1| |\vec p_4|
\delta_{m,1}-{p_1}_{j_1}{p_4}_{j_4}\delta_{m,0})\,, \lab{eq:Wjj}
\ee
in the basis of polarization vectors basis
(cf.~\cite{Bartels:2009sc}), can be written as
\ba \lab{eq:Wll} {\cal W}^{m_1}_{\lambda_1\lambda_4} (\vec
p_1,\vec p_4) &=& \sum_{j_1,j_4} \epsilon^{(\lambda_1)
}_{j_1}(\vec p_1) \epsilon^{(\lambda_4)}_{j_4}(\vec p_4)^{\ast}
W^{m_1}_{j_1 j_4}(\vec p_1,\vec p_4)
\nonumber \\[2mm] &\approx& |\vec p_1|
|\vec p_4| (\delta_{m_1,1}
\delta_{\lambda_1,h}\delta_{\lambda_4,h}+\delta_{m_1,0}
\delta_{\lambda_1,L}\delta_{\lambda_4,L}) \, . \ea
In analogy with~\cite{Bartels:2009sc}, we introduce the impact
factor for the coupling of two gravitons
\be \Phi_{\lambda_1\lambda_4}(|\vec p_1|,|\vec p|, |\vec
p_4|;z_0,y_0) \ = \ \sum_{m=0,1} {\cal W}^{m}_{\lambda_1\lambda_4}
(\vec p_1,\vec p_4) K_{m}(z_0|\vec p_1|) \tilde {\cal
K}_{m}(z_0,y_0;|\vec p|) K_{m}(y_0|\vec p_4|)\,.
\lab{eq:2gravimpact} \ee
We rewrite eq.\ (\ref{eq:dbT}) as
\ba T_{(14)\mu \nu, \rho \sigma}(z_0,y_0;\vec p_1,\vec p;\vec p_4)
&\approx& - \ft{1}{8} z_0^4 y_0^4 {p_1}_{k_1}
 p_{k_2}
p_{k_3}
 {p_4}_{k_4}
(\delta_{\mu k_1}\delta_{\nu k_3} +\delta_{\mu k_3}\delta_{\nu
k_1} ) \nonumber \\[2mm] &&(\delta_{\rho k_2}\delta_{\sigma k_4} +\delta_{\rho
k_4}\delta_{\sigma k_2}) \Phi_{\lambda_1\lambda_4}(|\vec
p_1|,|\vec p|, |\vec p_4|;z_0,y_0)\,. \ea
For the {\em lower} stress-energy tensors we make use of the
impact factors introduced in~\cite{Bartels:2009sc}
\be \Phi_{\lambda_2\lambda_5}(|\vec p_2|,|\vec p_5|;w_0) \ = \
\sum_{m=0,1} {\cal W}^{m}_{\lambda_2\lambda_5} (\vec p_2,\vec p_5)
K_m(w_0|\vec p_2|) K_m(w_0|\vec p_5|)\,. \lab{eq:1gravimpact} \ee
With this notation, the {\em lower} stress-energy tensors can be
written in the form
\be T_{(25)\mu' \nu'}(w_0;\vec p_2,\vec p_5) \ \approx\  2 w_0^4
{p_2}_{k_2'}
 {p_5}_{k_5'}
(\delta_{\mu' k_2'}\delta_{\nu' k_5'} +\delta_{\mu'
k_5'}\delta_{\nu' k_2'}) \, \Phi_{\lambda_2\lambda_5}(|\vec
p_2|,|\vec p_5|;w_0) \ee
and
\be T_{(36)\rho' \sigma'}(v_0;\vec p_3,\vec p_6) \ \approx\  2
v_0^4 {p_3}_{k_3'}
 {p_6}_{k_6'}
(\delta_{\rho' k_3'}\delta_{\sigma' k_6'} +\delta_{\rho'
k_6'}\delta_{\sigma' k_3'}) \, \Phi_{\lambda_3\lambda_6}(|\vec
p_3|,|\vec p_6|;v_0)\,. \ee
We note that, similarly to the four point correlators
in~\cite{Bartels:2009sc}, helicity is conserved in all impact
factors. With the vector $\vec p =\vec p_1+ \vec p_2+\vec p_5$
from eq.\ (\ref{eq:vectorp}) and with
\be M^2\ = \  -{\vec p}^2 \ee
we now perform the Wick rotation to positive $M^2$: $|\vec p| \to
iM$. In the limit of large $s_1$ and $s_2$ we thus arrive at:
\ba {\cal A}^{\rm 2G, planar}_{\lambda_1 \lambda_2 \lambda_3
\lambda_4 \lambda_5 \lambda_6} & = & 2 s_1^2 s_2^2 \int_0^{\infty}
d z_0 \int_0^{\infty}  d y_0 \int_0^{\infty} d w_0 \int_0^{\infty}
d v_0\,
 z_0^3 y_0^3
w_0^3 v_0^3\,  \Phi_{\lambda_1\lambda_4}(|\vec p_1|,i M, |\vec
p_4|;z_0,y_0)
\nonumber\\[2mm] && \hspace*{-1cm}
{\cal G}(z_0,w_0;\vec p_2+\vec p_5) {\cal G}(y_0,v_0;\vec p_3+\vec
p_6)\,  \Phi_{\lambda_2\lambda_5}(|\vec p_2|,|\vec p_5|;w_0)
\Phi_{\lambda_3\lambda_6}(|\vec p_3|,|\vec p_6|;v_0)\,.
\lab{eq:XPPP} \ea
This formula summarizes our results for the high energy limit of
the planar amplitude in fig.\ \ref{fig:gravboson}. The second
Witten diagram with crossed bulk-to-bulk graviton propagators can
now be obtained very easily. Introducing the vector
\be \vec p'\ = \ \vec p_1+ \vec p_3+\vec p_6\,, \ee with \be |\vec
p'|^2\ = \ M^2+t-t_1-t_2+|\vec p_1|^2+|\vec p_4|^2 \approx
M^2+|\vec p_1|^2+|\vec p_4|^2\,, \ee
the high energy limit of the crossed diagram has the form
\ba {\cal A}^{\rm 2G,crossed}_{\lambda_1 \lambda_2 \lambda_3
\lambda_4 \lambda_5 \lambda_6} &=& 2 s_1^2 s_2^2 \int_0^{\infty} d
z_0 \int_0^{\infty}  d y_0 \int_0^{\infty}  d w_0 \int_0^{\infty}
d v_0\,
 z_0^3 y_0^3
w_0^3 v_0^3 \, \Phi_{\lambda_1\lambda_4}(|\vec p_1|,|\vec p'|,
|\vec p_4|;z_0,y_0)
\nonumber\\[2mm] && \hspace*{-1cm}
{\cal G}(y_0,w_0;\vec p_2+\vec p_5) {\cal G}(z_0,v_0;\vec p_3+\vec
p_6)\, \Phi_{\lambda_2\lambda_5}(|\vec p_2|,|\vec p_5|;w_0)
\Phi_{\lambda_3\lambda_6}(|\vec p_3|,|\vec p_6|;v_0)\,.
\lab{eq:XPPPcr} \ea
For large $M^2$ we could substitute $|\vec p'| \to M$, but for the moment
we keep $M^2$ finite.

\subsection{Analytic structure of the two graviton impact factor}

In the last section we have identified the two graviton impact
factor (\ref{eq:2gravimpact}) as one of the new building blocks
for the planar amplitude. Let us pause for a moment and have a
closer look at its analytic structure.  We are interested in the
region where $M^2=-|\vec p|^2$ is positive and we have substituted
$|\vec p| \to - iM$. The impact factor contains the function
${\cal \tilde K}_m(z_0,y_0;|\vec p|)$ that arises from the
intermediate bulk-to-bulk $R$-boson propagator and is defined as
the analytic continuation of $\tilde {\cal K}_m(z_0,y_0;M)$. Since
$\tilde {\cal K}_m(z_0,y_0;M)$ is defined as an infinite sum over
modified Bessel functions, see eq.\ (\ref{eq:tKa}), its analytic
continuation
\be (\mp i M)^n K_n(\mp iM)\ = \ - \ft{\pi}{2} M^n (Y_n(M)\mp i
J_n(M))\,, \ee
has a cut for positive $M^2$ with a branching point at $M^2=0$,
its discontinuity being given by $\pi M^n J_n(M)$. While the upper
sign corresponds the region above the cut which is related to the
Feynman propagator, the lower sign is valid below the cut.

The analytic structure becomes more transparent if we make use of
another representation of the bulk-to-bulk $R$-boson propagator
\cite{Brower:2007qh,Avsar:2009xf}
\ba \tilde {\cal K}_{m}(z_0, y_0;|\vec p|) &=&
 \int_{0}^{\infty}
\frac{\omega \,d \omega }
{\omega^2+|\vec p|^2}
J_{m}(\omega z_0)
J_{m}(\omega y_0)
\nonumber \\[2mm]
&=&
K_{m}(z_0 |\vec p|)
I_{m}(y_0 |\vec p|)
\theta(z_0-y_0)
+
K_{m}(y_0|\vec p|)
I_{m}(z_0 |\vec p|)
\theta(y_0-z_0)\,,
\lab{eq:KIep}
\ea
where $K_a(x)$ and $I_a(x)$ are modified Bessel functions. The
subscripts $m=1$ and $m=0$ correspond to the transverse and
longitudinal polarization, respectively. Making use of the first
line on the right hand side of eq.\ (\ref{eq:KIep}), one can
rewrite the two graviton impact factor as a superposition of
products of single graviton impact factors
\ba \Phi_{\lambda_1\lambda_4}(|\vec p_1|,|\vec p|, |\vec
p_4|;z_0,y_0)&=& \int_{0}^{\infty} \frac{\omega \,d \omega }
{\omega^2+|\vec p|^2} \sum_{m=0,1} {\cal
W}^{m}_{\lambda_1\lambda_4} (\vec p_1,\vec p_4) K_{m}(z_0|\vec
p_1|) J_{m}(\omega z_0) \nonumber\\[2mm] && \hspace*{3cm} J_{m}(\omega y_0)
K_{m}(y_0|\vec p_4|)\,. \lab{eq:2gif} \ea
Using eq.\ (\ref{eq:Wll}), the second line on the right hand side
can be rewritten as
\ba \Phi_{\lambda_1\lambda_4}(|\vec p_1|,|\vec p|, |\vec
p_4|;z_0,y_0)&=& \frac{1}{|\vec p_1| |\vec p_4|} \int_{0}^{\infty}
\frac{\omega \,d \omega } {\omega^2+|\vec p|^2} \sum_{m=0,1} {\cal
W}^{m}_{\lambda_1\lambda_4} (\vec p_1,\vec p_4) K_{m}(z_0|\vec
p_1|) J_{m}(\omega z_0)\nonumber\\[2mm]&& \hspace*{2cm} \sum_{m'=0,1} {\cal
W}^{m'}_{\lambda_1\lambda_4} (\vec p_1,\vec p_4) J_{m'}(\omega
y_0) K_{m'}(y_0|\vec p_4|)\,. \ea
Performing the Wick rotation, substituting $|\vec p| \to iM$ and
comparing with the single graviton impact factor in eq.\
(\ref{eq:1gravimpact}) we identify the right hand side  as a
dispersion integral over the product of the imaginary parts of two
single graviton impact factors, where one of the currents has been
analytically continued into the time like region
\ba \Phi_{\lambda_1\lambda_4}(|\vec p_1|,- iM, |\vec
p_4|;z_0,y_0)&=& \frac{4}{\pi^2} \frac{1}{|\vec p_1| |\vec p_4|}
\int_{0}^{\infty} \frac{\omega \,d \omega } {\omega^2-M^2} \Im(i^m
\Phi_{\lambda_1\lambda_4}( |\vec p_1|,-i \omega;z_0) )\, \times
\nonumber\\[2mm]&& \hspace*{3cm}
\times \, \Im(i^m \Phi_{\lambda_1\lambda_4}(-i \omega, |\vec
p_4|;y_0)) \,. \ea
On the other hand the dispersion integral is given by
\ba \Phi_{\lambda_1\lambda_4}(|\vec p_1|,- iM, |\vec
p_4|;z_0,y_0)&=& \frac{1}{\pi} \int_{0}^{\infty} \frac{2 \omega
\,d \omega } {\omega^2-M^2} \Im( \Phi_{\lambda_1\lambda_4}(|\vec
p_1|,-i \omega, |\vec p_4|;z_0,y_0) )\,. \ea
Comparing the previous two equations we conclude  that the
imaginary part of the two graviton impact factor is equal to the
product of imaginary parts of two single graviton impact factors.

Finally, it is also interesting to investigate the behavior of the
two graviton impact factor for large values of $M^2$. Making use
of the integral representation (\ref{eq:KIep}) of the $R$-boson
propagator along with the completeness relation for Bessel
functions, one can expand the propagator for large $M$ to
obtain
\ba {\cal \tilde K}_{\epsilon_P}(z_0,y_0;,|\vec p|)\ = \
\frac{\delta(z_0-y_0)}{z_0 |\vec p|^2} -\frac{1}{|\vec p|^4}
\int_0^\infty d \omega \omega^3 J_{\epsilon_P}( \omega z_0 )
J_{\epsilon_P}( \omega y_0 ) +\ldots\,. \lab{eq:Kexp} \ea
A similar analysis also applies to the crossed amplitude. For
large $M^2$ we have $|\vec p'|^2\approx M^2$. Therefore the
leading contributions proportional to $1/M^2$ cancel from the sum
of the two diagrams. We are left with the asymptotic behavior
$\sim 1/M^4$ of the combined amplitude. This behavior of the two
graviton impact factor (\ref{eq:2gravimpact}) may be compared with
the analogous impact factor on the weak coupling side,
$\Phi^{direct}$ in eq.\ (\ref{eq:Fdirect}). It is curious to
observe that the latter has the same asymptotic behavior $\sim
1/M^4$ for large values of $M^2$.

\subsection{The deep inelastic limit}

In this subsection we turn to the diffractive cross section which
is given by the discontinuity of the six-point correlator across
the positive $M^2$ cut. For this discussion we specialize on the
kinematic limit where the virtualities of the {\em upper} bosons
are much larger than the virtualities of the {\em lower} ones,
namely
\ba |\vec p_1|^2, |\vec p_4|^2 \gg |\vec p_2|^2, |\vec p_3|^2,
|\vec p_5|^2, |\vec p_6|^2. \label{eq:DISlimit} \ea
For further simplification we set
\ba |\vec p_1|^2 \ = \ |\vec p_4|^2 \ = \ Q_A^2 \ea and \ba |\vec
p_2|^2\ = \  |\vec p_3|^2\ = \ |\vec p_5|^2\ = \  |\vec p_6|^2\ =
\  Q_B^2. \ea
This is the kinematic limit probed in, e.g., deep inelastic
electron proton scattering; for this reason we name this limit as
'deep inelastic limit'. This limit will allow us to perform the
integrations over the fifth coordinates and to obtain explicit
analytic expressions. In particular, this limit will allow us to
study the large-$M^2$ behavior of the imaginary part of the impact
factor which, in the diffractive cross section, determines the
large-$M^2$ behavior of the cross section.

To simplify notation we define dimensionless variables
\be z_M\ = \ z_0 M \,,\quad y_M\ = \ y_0 M \,,\quad v_M\ = \ v_0 M
\,,\quad w_M\ = \ w_0 M \,, \ee the ratios \be \alpha\ = \ Q_A
/M\,, \quad \beta\ = \ Q_B/Q_A\,. \ee and \be \varepsilon_k\ = \
|\vec q_k|/Q_B \,,\quad \vec q_1\ = \ \vec p_2+\vec p_5 \,,\quad
\vec q_2\ = \ \vec p_3+\vec p_6\,. \ee
With these definitions we rewrite the the planar amplitude
(\ref{eq:XPPP}) as
\ba {\cal A}^{\rm 2G,planar}_{\lambda_A \lambda_{B1}\lambda_{B2}}
&=& 2 \left(\frac{s_1}{Q_A^2}\right)^2
\left(\frac{s_2}{Q_A^2}\right)^2 \alpha^{16} \beta^4 Q_A^{-2}
\int_0^{\infty}  d y_M \int_0^{\infty}  d v_M \int_0^{\infty}  d
w_M \int_0^{\infty}  d z_M
 w_M^3 v_M^3
z_M^3
y_M^3
\nonumber \\[2mm] &&
K_{m(\lambda_A)}(z_M \alpha)
{\cal \tilde K}_{m(\lambda_A)}(z_M,y_M; -i )K_{m(\lambda_A)}(y_M \alpha)
\nonumber\\[2mm]
&&
{\cal G}(z_M,w_M;\varepsilon_1 \alpha \beta)
{\cal G}(y_M,v_M;\varepsilon_2 \alpha \beta)
\nonumber \\[2mm]
&&
K_{m(\lambda_{B1})}(w_M \alpha \beta)
K_{m(\lambda_{B1})}(w_M \alpha \beta)
K_{m(\lambda_{B2})}(v_M \alpha \beta)
K_{m(\lambda_{B2})}(v_M \alpha \beta) \,.
\ea
Here we have inserted the definitions of the impact factors.
Making use use of helicity conservation we can rename the helicity
variables such  that $\lambda_A=\lambda_1=\lambda_4$ and
$\lambda_{B1}=\lambda_2=\lambda_5$,
$\lambda_{B2}=\lambda_3=\lambda_6$. Furthermore, $m(\lambda)=0$
for longitudinal polarization, and $m(\lambda)=1$ for transverse
polarization. We have also a similar expression for the crossed
diagram.

As a first step of simplification let us consider the forward
limit
\be \varepsilon_k \to 0\,, \ee
for $k=1,2$, i.e. $t_1=t_2\to 0$, so that the graviton propagator
\be {\cal G}(z_M,w_M;0) \ = \ \frac{1}{4} ( \frac{w_M^2}{z_M^2}
\theta(z_M-w_M) +\frac{z_M^2}{w_M^2} \theta(w_M-z_M) )\,. \ee
Then
\ba {\cal A}^{\rm 2G,planar}_{\lambda_A \lambda_{B1}\lambda_{B2}}
&=& \frac1{8} \left(\frac{s_1}{Q_A^2}\right)^2
\left(\frac{s_2}{Q_A^2}\right)^2 \alpha^{16} \beta^4 Q_A^{-2}
\int_0^{\infty}  d y_M \int_0^{\infty}  d v_M \int_0^{\infty}  d
w_M \int_0^{\infty}  d z_M
 w_M^3 v_M^3
z_M^3
y_M^3
\nonumber \\[2mm] &&
K_{m(\lambda_A)}(z_M \alpha)
{\cal \tilde K}_{m(\lambda_A)}(z_M,y_M;- i )K_{m(\lambda_A)}(y_M \alpha)
\nonumber \\[2mm] &&
\left(\frac{w_M^2}{z_M^2} \theta(z_M-w_M)\frac{v_M^2}{y_M^2} \theta(y_M-v_M)
+\frac{z_M^2}{w_M^2} \theta(w_M-z_M)\frac{v_M^2}{y_M^2} \theta(y_M-v_M)
\right.
\nonumber \\[2mm] &&
\left.
+\frac{w_M^2}{z_M^2} \theta(z_M-w_M)\frac{y_M^2}{v_M^2} \theta(v_M-y_M)
+\frac{z_M^2}{w_M^2} \theta(w_M-z_M)\frac{y_M^2}{v_M^2} \theta(v_M-y_M)
\right)
\nonumber \\[2mm] &&
K_{m(\lambda_{B1})}(w_M \alpha \beta)
K_{m(\lambda_{B1})}(w_M \alpha \beta)
K_{m(\lambda_{B2})}(v_M \alpha \beta)
K_{m(\lambda_{B2})}(v_M \alpha \beta) \,.
\ea
Making use of expressions given in the Appendix \ref{sc:int} is is
possible to do the integrals over $w_M$ and $v_M$, and with the
saddle point method described in  Appendix \ref{sc:alt}, one can
investigate the large $M^2$ limit. However, we chose another way.

We turn to the DIS limit (\ref{eq:DISlimit}), which implies $\beta
\to 0$, and we expand in powers of $\beta$. Due to the fast
vanishing of the Bessel functions of the two graviton vertex
(which do not contain the $\beta$ variable) one can the {\it
lower} impact factors in powers of $\beta$ and perform the
integrals over $w_M$ and $v_M$. In the case of transverse
polarizations of the lower $R$-currents, the small-$\beta$
behavior of the Bessel functions gives rise to logarithmic
divergences for small $\beta$. The appearance of such logarithms
is known already from the single graviton exchange
~\cite{Bartels:2009sc}. For two gravitons we have maximally two
logarithms in $\beta$. Using eq.\ (\ref{eq:KIep}) one can then
perform the integrals over $z_M$ and $y_M$. Thus, the amplitude of
the planar diagram becomes
\ba {\cal A}^{\rm 2G,planar}_{\lambda_A \lambda_{B1}\lambda_{B2}}
&\approx& - M^{-2}\left(\frac{s_1}{Q_A^2}\right)^2
\left(\frac{s_2}{Q_A^2}\right)^2 I_{\lambda}(-\alpha^2)\,
\log^{m(\lambda_{B1})+m(\lambda_{B2})}(\beta^{-2})\,,
\lab{eq:Aplan} \ea
where the function $I_{\lambda}(-\alpha^2)$ stands for the result
of the integrals over $z_M$ and $y_M$
\ba I_{\lambda}(- \alpha^2) &=&- \frac{\alpha^{-2}}{32}
\int_0^{\infty} d z_M \int_0^{\infty} d y_M z_M^5 y_M^5
K_{m(\lambda)}(z_M) \tilde {\cal
K}_{m(\lambda)}(z_M/\alpha,y_M/\alpha;-i) K_{m(\lambda)}(y_M) \,.
\nonumber \\[2mm]
\lab{eq:Ipra} \ea
The integrations can be done analytically leading to
\be - \alpha^2 I_\lambda(-\alpha^2)\ = \ \left(p^{(0)}_\lambda +
p^{(1)}_\lambda \log(-\alpha^2) + p^{(2)}_\lambda
\log(\rho)\right)_{\rho=1}\,. \lab{eq:Ilambda} \ee
The functions $p^{(i)}_\lambda$ are rational functions in $\alpha$
and $\rho$, and their detailed form is presented in Appendix B
\footnote{In the appendix we discuss the more general case $|p_1|
\neq |p_4|$ and consider the function $I_{\lambda}$ as a function
of $\alpha$ and $\rho=|p_1|/|p_4|$. The results of this section
are obtained by taking the limit $\rho=1$.}. Due to the
$\ln(-\alpha^2)$, the function $I_{\lambda}$ has a cut for real
positive $\alpha^2= Q_A^2/M^2$, i.e a right cut in $M^2$ starting
at $M^2=0$. There no no poles in $M^2$. If we would have taken the
virtualities of the currents $\vec p_1$ and $\vec p_2$ to be
different from each other, we would have obtained also logarithms
in the ratio $\vec p_1/\vec p_2$. For further details we refer to
Appendix C.

The contribution related to the crossed diagram is obtained by
substituting: $-M^2 \to \tilde M^2\equiv M^2+t-t_1-t_2+|\vec
p_1|^2+|\vec p_4|^2$, i.e.\ ${\cal A}^{\rm 2G,crossed}_{\lambda_A
\lambda_{B1}\lambda_{B2}}$ is obtained from the analytic
continuation of ${\cal A}^{\rm 2G,planar}_{\lambda_A
\lambda_{B1}\lambda_{B2}}$ in the $M^2$ plane. As we have already
discussed before, in the large-$M^2$ limit the leading term of
${\cal A}^{\rm 2G,crossed}_{\lambda_A \lambda_{B1}\lambda_{B2}}$,
is of the order $\tilde M^{-2} \approx - M^{-2}$, and it cancels
with the leading term of ${\cal A}^{\rm 2G,planar}_ {\lambda_A
\lambda_{B1}\lambda_{B2}}$. This means that the sum is of the
order $M^{-4}$,
\be {\cal A}^{\rm 2G,planar}_{\lambda_A \lambda_{B1}\lambda_{B2}}
+ {\cal A}^{\rm 2G,crossed}_{\lambda_A \lambda_{B1}\lambda_{B2}} \
=\  - \frac{Q_A^2}{M^{4}}\left(\frac{s_1}{Q_A^2}\right)^2
\left(\frac{s_2}{Q_A^2}\right)^2 \hat I_{\lambda}(\alpha^2)
\log^{m(\lambda_A)+m(\lambda_B)}(\beta^{-2})\,. \nonumber
\lab{eq:Atot} \ee
\begin{figure}
\begin{center}
{
\psfrag{'laT.dat'}{$\lambda=T$}
\psfrag{'laL.dat'}{$\lambda=L$}
\psfrag{al}{$M^2/Q_A^2$}
\epsfysize7.0cm \epsfbox{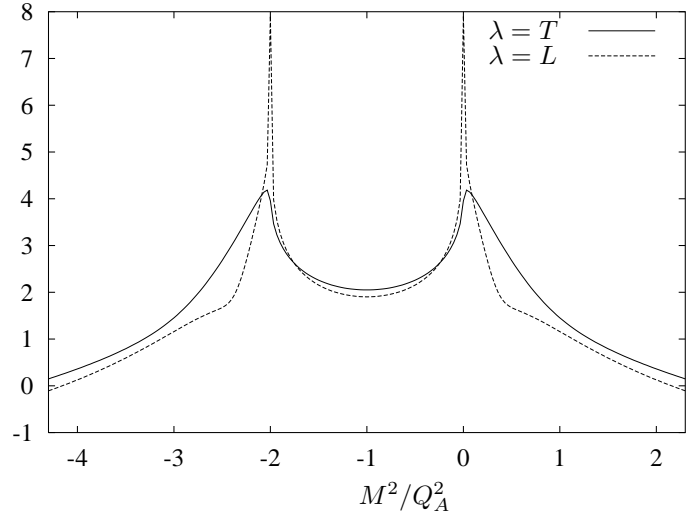}}
\end{center}
\caption{The logarithm of the absolute value
of
$\alpha^4 \hat I_{\lambda}(\alpha^2)$
plotted as a function of
$\alpha^{-2}=M^2/Q_A^2$.}
\lab{fig:hIa}
\end{figure}
\begin{figure}
\begin{center}
{
\psfrag{'puT.dat'}{$\lambda=T$}
\psfrag{'puL.dat'}{$\lambda=L$}
\psfrag{al}{$M^2/Q_A^2$}
\epsfysize7.0cm \epsfbox{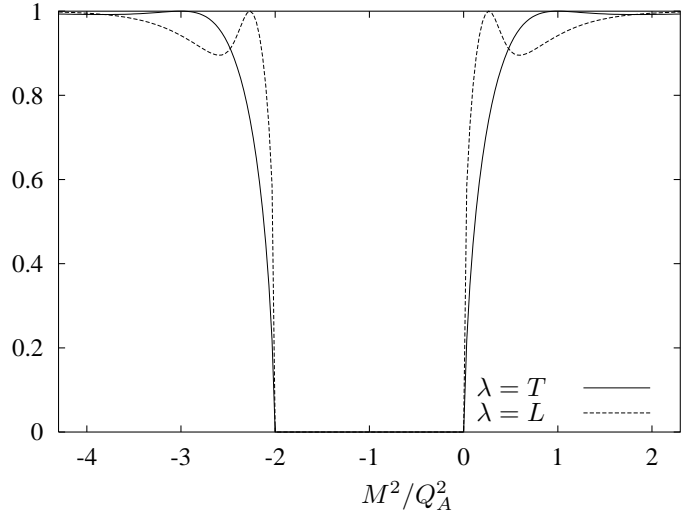}}
\end{center}
\caption{
The phase of the {\em upper} impact factor,
namely
$\pi^{-1} \arg(- \alpha^4 \hat I_{\lambda}(\alpha^2))$,
plotted as a function of
$\alpha^{-2}=M^2/Q_A^2$.
For $M^2>0$ we present the branch above the cut
while for $M^2<0$ we show the phase below the cut.
}
\lab{fig:hIp}
\end{figure}
The function
\be \hat I_{\lambda}(\alpha^2) \ = \
 \alpha^{-2}
(I_{\lambda}(- \alpha^2) - I_{\lambda}((\alpha^{-2}+2)^{-1}))\,,
\ee
describing the sum of the planar and crossed impact factor has
both right and left hand cuts in $M^2$. The absolute value of
$\alpha^4 \hat I_{\lambda}(\alpha^2)$ is shown fig.~\ref{fig:hIa}
and \ref{fig:hIp}, both for transverse and longitudinal
polarizations. In both cases, there is a maximum at the beginning
of the $M^2$-cuts. In contrast to the transverse impact factor,
the longitudinal one is logarithmically divergent at $M^2=0$ and
$\tilde M^2=0$. These divergences come from the logarithmic
behavior of the longitudinal $R$-boson propagator (\ref{eq:tKa}).
In the large $M^2$ limit the leading term of $\hat
I_{\lambda}(\alpha^2=0)$ is of the form
\be \hat I_{\lambda}(\alpha^2=0) \ = \ \int_0^{\infty}  dr r^7
K_{m(\lambda)}^2(r) \ = \  \frac{8}{35} \Gamma(4-m(\lambda))
\Gamma(4+m(\lambda)) \,. \ee

From eq.\ (\ref{eq:Ilambda}), with the explicit form of
$p^{(1)}_{\lambda}$ being given in the appendix, it is
straightforward to determine the discontinuity of the amplitude
(\ref{eq:Aplan}) across the right hand cut in $M^2$
\ba {\rm disc}_{M^2} I_{T}(-\alpha^2)&=& - \frac{576 \alpha ^{12}
\left(\alpha ^2-1\right) \pi \left(\alpha ^2
-1\right)}{\left(\alpha
 ^2+1\right)^5 \left(\alpha ^2 +1\right)^5}\,,
\lab{eq:iIT}
\ea
and
\ba {\rm disc}_{M^2} I_{L}(-\alpha^2)&=& - \frac{64 \alpha ^{10}
\left(\alpha ^4-4 \alpha
 ^2+1\right) \pi  \left(\alpha ^4 -4
 \alpha ^2 +1\right)}{\left(\alpha ^2+1\right)^5
 \left(\alpha ^2 +1\right)^5}\,.
\lab{eq:iIL}
\ea
For $M \to 0$ the discontinuity for transversely polarized $R-$bosons
vanishes as $M^2$,
while the longitudinally polarized one goes to a constant.
For large $M^2$, the imaginary part of
${\cal A}^{\rm 2G,planar}_{\lambda_A \lambda_{B1}\lambda_{B2}}
$
is proportional to $M^{-12}$ ($M^{-14}$) for the longitudinal
(transverse) polarization.
Finally, one can also notice that the rescaled imaginary part,
$\alpha^{-4} \Im I_{\lambda}(-\alpha^2)$,
is invariant under the substitution
$\alpha^2 \leftrightarrow (\alpha^2 )^{-1}$.

We end this section with a comment on the diagram on the rhs of fig.4. It contains 
a direct coupling of two gravitons to the upper $R$ boson, and it does 
not depend upon $M^2$. At the end of the following section we will show 
that its dependence upon $s_1$ and $s_2$ is quite similar to the 
triple graviton diagram to which we turn in the following section.     

\section{The triple Regge limit: triple graviton exchange}
\lab{sc:TGV}

There are two more Witten diagrams that can contribute to the
six-point correlators of R-currents, namely the two terms that are
depicted in fig. \ref{fig:grav3}. The first one involves the
triple graviton vertex. We will construct the vertex in the
following subsection before we evaluate the Regge limit of the
entire diagram in subsection 4.2. The second diagram in fig.
\ref{fig:grav3} is the subject of subsection 4.3. It contains a
vertex between two gravitons and two R-bosons. Through our
analysis, the only term that could contribute to the discontinuity
in $M^2$ is found to vanish. Furthermore, we shall show that the
remaining $M^2$-independent terms from the two diagrams in fig.
\ref{fig:grav3} are subleading compared to the contributions from
the Witten diagrams in fig.\ \ref{fig:gravboson}.

\subsection{Triple graviton vertex}

In order to analyze the first diagram of fig. \ref{fig:grav3} we
need an expression for the three graviton vertex. 
This vertex was derived before in Ref.~\cite{Arutyunov:1999nw}.
In the following, we re-derive the vertex at prepare for the high energy limit.
As usual, our task is to expand the Einstein
Hilbert action
\be S\ = \ -\frac1{2 \kappa_{d+1}^2}\int d^{d+1}z \sqrt{g} R\,,
\ee
in small fluctuations $h_{\mu \nu}$ of the metric $ g_{\mu\nu} \ =
\ \bar g_{\mu \nu}+ h_{\mu \nu}$ around the metric $\bar g_{\mu
\nu}$ of the AdS background. In order to fix our conventions we
recall that the curvature, Ricci tensor and Riemann tensor are
defined through
\ba R&=&R_{\mu \nu} g^{\mu \nu} \ = \ R_{\mu \alpha \beta \nu }
g^{\mu \nu} g^{\alpha \beta}\,, \label{eq:Rdef} \ea
\ba R_{\mu \alpha \beta \nu}&=& g_{\mu \gamma} R^{\gamma}_{\;
\alpha \beta \nu} \ = \ g_{\mu \gamma} (
\partial_{\beta} \Gamma^{\gamma}_{\alpha \nu}
-\partial_{\nu} \Gamma^{\gamma}_{\alpha \beta}
+\Gamma^{\gamma}_{\rho \beta} \Gamma^{\rho}_{\alpha \nu}
-\Gamma^{\gamma}_{\rho \nu} \Gamma^{\rho}_{\alpha \beta} )\,, \ea
where the Christoffel symbols are given by
 \ba
\Gamma^{\alpha}_{\beta \gamma}\ = \ \ft1{2} g^{\alpha \rho} (
\partial_\beta g_{\gamma \rho}
+\partial_\gamma g_{\beta \rho} -\partial_\rho g_{\beta \gamma}
)\,. \ea
In the following calculation we need to expand both the inverse
metric $g^{\alpha \beta}$ and the determinant $g$ up to third
order in the fluctuation $h_{\mu\nu}$. For the inverse metric we
find
\be g^{\alpha \beta} \, = \, \bar g^{\alpha \beta} -\bar g^{\alpha
\mu_1} h_{\mu_1 \nu_1} \bar g^{\nu_1 \beta} +\bar g^{\alpha \mu_1}
h_{\mu_1 \nu_1} \bar g^{\nu_1 \mu_2} h_{\mu_2 \nu_2} \bar g^{\nu_2
\beta} -\bar g^{\alpha \mu_1} h_{\mu_1 \nu_1} \bar g^{\nu_1 \mu_2}
h_{\mu_2 \nu_2} \bar g^{\nu_2 \mu_3}h_{\mu_3 \nu_3} \bar g^{\nu_3
\beta} \ldots\,, \nonumber \ee
while
\ba \sqrt{g}&=& \exp{\ln \sqrt{g}} \ \approx\ \sqrt{\bar g} (1 +
\ft{1}{2} \bar g^{ \sigma \rho} h_{\rho \sigma} - \ft1{4} \bar g^{
\sigma \rho} h_{\rho \nu_1} \bar g^{ \nu_1 \rho_1} h_{\rho_1
\sigma} + \ft1{8} (\bar g^{ \sigma \rho} h_{\rho \sigma})^2 +
\nonumber\\[2mm] & & \hspace*{-1cm} + \ft1{6}
  \bar g^{ \sigma \rho} h_{\rho \nu_1}
 \bar g^{ \nu_1 \rho_1} h_{\rho_1 \nu_2}
 \bar g^{ \nu_2 \rho_2} h_{\rho_2 \sigma}
 - \ft1{8} ( \bar g^{ \sigma \rho} h_{\rho \nu_1}
 \bar g^{ \nu_1 \rho_1} h_{\rho_1 \sigma})
 \bar g^{ \nu_3 \rho_3} h_{\rho_3 \nu_3}
+ \ft{1}{48} (\bar g^{ \sigma \rho} h_{\rho \sigma})^3 ).
\lab{eq:sqg} \ea
After the substitution $ g_{\mu \nu } \to \bar g_{\mu \nu }+
h_{\mu \nu }$ we can expand the Langrangian of the Einstein
Hilbert action,
\ba -\sqrt{g} R\ = \ -\sqrt{\bar g} \left(\bar R+
 H^{(1)}
+ H^{(2)} + H^{(3)} \right)\,, \ea
up to third order in the fluctuation field $h_{\mu\nu}$. The
constant term is determined by the AdS curvature $-\bar R\ = \
-d(d+1)$. The first order corrections to the curvature $\bar R$
involve the quantity
\ba
 H^{(1)}&=&
- z_0^2 (d-2) (d-1)
h_{00}(z)
+\frac{1}{2} z_0^2 ((d-3) d+4)
\bar h
- z_0^3 (d-4)
\partial_0 \bar h
\nonumber \\[2mm] & & \hspace*{2cm}
+ z_0^3 \, 2 (d-2)
\partial_\alpha h_{\alpha 0}
+ z_0^4(
\partial_\alpha\partial_\alpha \bar h
-
\partial_\alpha\partial_\beta h_{\alpha \beta}
)\,,
\ea
where $\bar h=h_{\alpha \alpha}$ is the trace of the fluctuation
field. After multiplication with the factor $\sqrt{g}$, we can
write these terms as a total derivative, in agreement with the
fact that we are expanding around a solution $\bar g$ of the
Einstein Hilbert action.
The equation of motion for the fluctuation field $h$ is related to
the second order terms $H^{(2)}$ in the expansion of the
Lagrangian. We have reproduced an explicit expression in Appendix
\ref{sc:va}. What we really need here is the form of the terms
that appear in the third order,
\ba
 H^{(3)}&=&
\frac{1}{48} ( (d-11) d
+36) z_0^6
{\bar h}^3
-\frac{1}{8} z_0^6 ((d-11) d+34)
h_{00} {\bar h}^2
+\frac{1}{2}(d -11) d z_0^6
\bar h h_{\alpha 0} h_{\alpha 0}
\nonumber \\[2mm] &&
+ \frac{1}{8} ((11- d )d -40)z_0^6
h_{\alpha \beta} h_{\alpha \beta} {\bar h}
+\frac{1}{4} z_0^6 ((d-11) d+38)
h_{\alpha \beta} h_{\alpha \beta} h_{00}
\nonumber \\[2mm] &&
- z_0^6 (d-8) (d-3)
h_{\alpha \beta} h_{\alpha 0} h_{\beta 0}
+\frac1{6}((d-11)d+48) z_0^6
h_{\alpha \beta}
h_{\alpha \gamma}
h_{\beta \gamma}
+\frac{1}{2} z_0^8
{\bar h} \partial_{\alpha} h_{\alpha \beta} \partial_{\gamma} h_{\beta \gamma}
\nonumber \\[2mm] &&
+15 z_0^6
h_{\alpha 0} h_{\alpha 0} {\bar h}
+9 z_0^7
h_{0 \gamma}
h_{\alpha \beta}
\partial_{\gamma} h_{\alpha \beta}
-
\frac{1}{8} z_0^7 (d-8)
{\bar h}^2 \partial_0 {\bar h}
+\frac{1}{8} z_0^8
\partial_{\alpha}\partial_{\beta} h_{\alpha \beta}(2 h_{\gamma \rho} h_{\gamma \rho}-{\bar h}^2)
\nonumber \\[2mm] &&
+\frac{1}{4} z_0^7 (d-6)
{\bar h}^2 \partial_\alpha h_{\alpha 0}
+
\frac{1}{2} z_0^7 (d-8)
{\bar h} h_{0 \alpha} \partial_\alpha {\bar h}
- z_0^7 (d-6)
{\bar h} h_{0 \alpha} \partial_\beta h_{\alpha \beta}
\nonumber \\[2mm] &&
+\frac{1}{2} z_0^7 (d-9)
{\bar h} h_{\alpha \beta} \partial_0 h_{\alpha \beta}
-z_0^7 (d-5)
{\bar h} h_{\alpha \beta} \partial_\alpha h_{\beta 0}
- z_0^7 d
h_{0 \gamma} h_{\beta \alpha} \partial_\gamma h_{\beta \alpha}
\nonumber \\[2mm] &&
- z_0^7 (d-8)
h_{0 \alpha} h_{\beta \alpha} \partial_\beta {\bar h}
+2 z_0^7 (d-6)
h_{0 \alpha} h_{\beta \alpha} \partial_\gamma h_{\beta \gamma}
+2 z_0^7 (d-5)
h_{0 \alpha} h_{\beta \gamma} \partial_\gamma h_{\beta \alpha}
\nonumber \\[2mm] &&
+ z_0^7 (d+2)
h_{\alpha \beta} h_{\alpha \gamma} \partial_0 h_{\beta \gamma}
+\frac{1}{4} z_0^7 (d-8)
h_{\alpha \beta} h_{\alpha \beta} \partial_0 {\bar h}
- \frac1{2} z_0^7 (d-6)
h_{\alpha \beta} h_{\alpha \beta} \partial_\gamma h_{\gamma 0}
\nonumber \\[2mm] &&
+
 z_0^8
h_{\beta \alpha} \partial_{\rho} h_{\alpha \rho} \partial_{\beta} h_{\gamma \gamma}
+\frac{3}{2} z_0^8
h_{\beta \alpha} \partial_{\gamma} h_{\beta \rho} \, \partial_{\gamma} h_{\alpha \rho}
- z_0^8
h_{\beta \alpha} \partial_{\beta} h_{\gamma \rho} \, \partial_{\gamma} h_{\alpha \rho}
-\frac{1}{2} z_0^8
h_{\beta \alpha} \partial_{\gamma} h_{\beta \rho} \, \partial_{\rho} h_{\alpha \gamma}
\nonumber \\[2mm] &&
- \frac{1}{2} z_0^8
h_{\beta \alpha} \partial_{\gamma} h_{\alpha \beta } \, \partial_{\gamma} {\bar h}
- \frac{1}{4} z_0^8
h_{\beta \alpha} \partial_{\alpha} {\bar h} \, \partial_{\beta} {\bar h}
- z_0^8
h_{\beta \alpha} h_{\rho \gamma} \partial_{\gamma} \partial_{\beta} h_{\alpha \rho}
+ \frac{3}{4} z_0^8
h_{\beta \alpha} \partial_{\beta} h_{\gamma \rho} \, \partial_{\alpha} h_{\rho \gamma}
\nonumber \\[2mm] &&
+ z_0^8
h_{\beta \alpha} \partial_{\beta} h_{\alpha \gamma} \, \partial_{\gamma} {\bar h}
-2 z_0^8
h_{\beta \alpha} \partial_{\beta} h_{\alpha \gamma} \, \partial_{\rho} h_{\gamma \rho}
+ z_0^8
h_{\beta \alpha} h_{\alpha \gamma} \, \partial_{\beta} \partial_{\gamma} {\bar h}
-2 z_0^8
h_{\beta \alpha} h_{\alpha \gamma} \,\partial_{\gamma} \partial_{\rho} h_{\beta \rho}
\nonumber \\[2mm] &&
+ z_0^8
h_{\beta \alpha} h_{\alpha \gamma} \,\partial_{\rho} \partial_{\rho} h_{\beta \gamma}
+ z_0^8
h_{\beta \alpha} \partial_{\gamma} h_{\alpha \beta} \,\partial_{\rho} h_{\gamma \rho}
+ z_0^8
h_{\beta \alpha} \partial_{\gamma} h_{\alpha \gamma} \,\partial_{\beta} {\bar h}
- z_0^8
h_{\beta \alpha} \partial_{\gamma} h_{\alpha \gamma} \,\partial_{\rho} h_{\beta \rho}
\nonumber \\[2mm] &&
+ z_0^8
{\bar h} h_{\beta \gamma} \partial_{\beta} \partial_{\alpha} h_{\alpha \gamma}
-\frac{1}{2} z_0^8
{\bar h} h_{\beta \gamma} \partial_{\alpha} \partial_{\alpha} h_{\beta \gamma}
-\frac{1}{2} z_0^8
{\bar h} \partial_{\alpha} h_{\alpha \beta } \partial_{\beta} {\bar h}
-\frac{1}{8} z_0^8
\partial_{\alpha}\partial_{\alpha} {\bar h}(2 h_{\gamma \rho} h_{\gamma \rho}-{\bar h}^2)
\nonumber \\[2mm] &&
-\frac{3}{8} z_0^8
{\bar h} \partial_{\gamma} h_{\alpha \beta} \partial_{\gamma} h_{\alpha \beta}
+\frac{1}{4} z_0^8
{\bar h} \partial_{\alpha} h_{\beta \gamma} \partial_{\beta} h_{\alpha \gamma}
+\frac{1}{8} z_0^8
{\bar h} \alpha {\bar h} \alpha {\bar h}
 -\frac{1}{2} z_0^8
{\bar h} h_{\beta \gamma} \partial_{\beta} \partial_{\gamma} {\bar h}
\,. 
\ea
In the following analysis we shall now specialize to $d=4$. Having
spelled out the third order terms $H^{(3)}$, we can now read off
the triple graviton vertex $V^{TR}$. In order to spell out the
answer, we shall split the vertex into four different
contributions,
\ba V^{TR}_{\mu_1 \nu_1,\mu_2\nu_2,\mu_3 \nu_3}(Q_1,Q_2,Q_3)&=&
V^{TR,11}_{\mu_1 \nu_1,\mu_2\nu_2,\mu_3 \nu_3}(Q_1,Q_2,Q_3)
+V^{TR,20}_{\mu_1 \nu_1,\mu_2\nu_2,\mu_3 \nu_3}(Q_1,Q_2,Q_3)
\nonumber \\[2mm] && \hspace*{-1cm} +\, V^{TR,10}_{\mu_1 \nu_1,\mu_2\nu_2,\mu_3
\nu_3}(Q_1,Q_2,Q_3) +V^{TR,00}_{\mu_1 \nu_1,\mu_2\nu_2,\mu_3
\nu_3}(Q_1,Q_2,Q_3)\,. \ea
Here, we group terms according to the number of the Kronecker
deltas which connect different gravitons, i.e.\ Kronecker deltas
of the form $\delta_{\mu_i, \nu_i}$ and those involving internal
(summed) labels are not counted. Explicitly, the terms that
contribute to $V^{TR,20}$ are given by
\ba
V&\!\!\!\!\!\!^{TR,20}_{\mu_1 \nu_1,\mu_2\nu_2,\mu_3 \nu_3}&\!\!\!\!\!\!(Q_1,Q_2,Q_3) \nonumber \\[2mm]
&=&
-\frac{3}{8} \delta_{\mu_1,\mu_2} \delta_{\nu_1,\nu_2}\delta_{\mu_3,\nu_3} Q_{1,\nu} Q_{2,\nu} z_0^3
-\frac{1}{2} \delta_{\mu_1,\mu_2} \delta_{\nu_1,\nu_2} \delta_{\mu_3,\nu_3} Q_{2,\nu} Q_{2,\nu} z_0^3
\nonumber \\[2mm] &&
+\frac{3}{4} \delta_{\mu_2,\nu_3} \delta_{\nu_2,\mu_3} Q_{2,\mu_1} Q_{3,\nu_1} z_0^3
+\delta_{\mu_1,\nu_2} \delta_{\nu_1,\mu_2} Q_{2,\mu_3} Q_{3,\nu_3} z_0^3
\nonumber \\[2mm] &&
-\frac{1}{2} \delta_{\mu_1,\nu_2} \delta_{\nu_1,\mu_2} \delta_{\mu_3,\nu_3} Q_{2,\nu} Q_{3,\nu} z_0^3
+\delta_{\mu_1,\nu_3} \delta_{\nu_1,\mu_3} Q_{3,\mu_2} Q_{3,\nu_2} z_0^3
\nonumber \\[2mm] &&
+\frac{1}{4} \delta_{\mu_1,\mu_2} \delta_{\nu_1,\nu_2} Q_{3,\mu_3} Q_{3,\nu_3} z_0^3
-\frac{1}{4} \delta_{\mu_1,\mu_2} \delta_{\nu_1,\nu_2} \delta_{\mu_3,\nu_3} Q_{3,\nu} Q_{3,\nu} z_0^3
\nonumber \\[2mm] &&
-\frac{5}{2} \delta_{\mu_1,\mu_2} \delta_{\nu_1,\nu_2} \delta_{\mu_3,\nu_3} Q_{2,0} z_0^2
+5 \delta_{\mu_2,\mu_3} \delta_{\nu_2,\nu_3} \delta_{0,\mu_1} Q_{3,\nu_1} z_0^2
\nonumber \\[2mm] &&
+\delta_{\mu_1,\mu_2} \delta_{\nu_1,\nu_2} \delta_{0,\nu_3} Q_{3,\mu_3} z_0^2
-\delta_{\mu_1,\mu_2} \delta_{\nu_1,\nu_2} \delta_{\mu_3,\nu_3} Q_{3,0} z_0^2
\nonumber \\[2mm] &&
-\frac{3}{2} \delta_{\mu_1,\mu_2} \delta_{\nu_1,\nu_2}
\delta_{\mu_3,\nu_3}z_0 +\frac{5}{2} \delta_{\mu_2,\mu_3}
\delta_{\nu_2,\nu_3} \delta_{0,\mu_1} \delta_{0,\nu_1} z_0\, .
\label{eq:VTR20}\ea
All terms we displayed contract the indices among two of the three
fluctuation fields. Terms in which the contractions involve all
three graviton fields are collected in
\ba V&\!\!\!\!\!\!^{TR,11}_{\mu_1 \nu_1,\mu_2\nu_2,\mu_3
\nu_3}&\!\!\!\!\!\!(Q_1,Q_2,Q_3)
\nonumber \\[2mm]
&=&
- \delta_{\nu_1,\mu_3} \delta_{\nu_2,\nu_3} Q_{2,\mu_1} Q_{3,\mu_2} z_0^3
-\frac{1}{2} \delta_{\mu_1,\mu_2} \delta_{\nu_1,\mu_3} Q_{2,\nu_3} Q_{3,\nu_2} z_0^3
\nonumber \\[2mm] &&
-2 \delta_{\nu_1,\mu_2} \delta_{\nu_2,\mu_3} Q_{2,\mu_1} Q_{3,\nu_3} z_0^3
-\delta_{\mu_1,\mu_3} \delta_{\nu_1,\mu_2} Q_{2,\nu_2} Q_{3,\nu_3} z_0^3
\nonumber \\[2mm] &&
+\frac{3}{2} \delta_{\mu_1,\mu_2} \delta_{\nu_1,\mu_3} \delta_{\nu_2,\nu_3} Q_{2,\nu} Q_{3,\nu} z_0^3
- \delta_{\nu_1,\mu_3} \delta_{\mu_2,\nu_3} Q_{3,\mu_1} Q_{3,\nu_2} z_0^3
\nonumber \\[2mm] &&
-2 \delta_{\mu_1,\mu_3} \delta_{\nu_1,\mu_2} Q_{3,\nu_2} Q_{3,\nu_3} z_0^3
+\delta_{\mu_1 ,\mu_3} \delta_{\nu_1,\mu_2} \delta_{\nu_2,\nu_3} Q_{3,\nu} Q_{3,\nu} z_0^3
\nonumber \\[2mm] &&
-2 \delta_{\nu_1,\nu_3} \delta_{\mu_2,\mu_3} \delta_{0,\mu_1} Q_{3,\nu_2} z_0^2
-4 \delta_{\nu_1,\nu_2} \delta_{\mu_2,\mu_3} \delta_{0,\mu_1} Q_{3,\nu_3} z_0^2
\nonumber \\[2mm] &&
+6 \delta_{\mu_1,\mu_3} \delta_{\nu_1,\mu_2} \delta_{\nu_2,\nu_3} Q_{3,0}z_0^2
+\frac{10}{3} \delta_{\mu_1,\mu_2} \delta_{\nu_1,\nu_3} \delta_{\nu_2,\mu_3} z_0
\nonumber \\[2mm] &&
+4 \delta_{\mu_1,\mu_2}
\delta_{\nu_1,\mu_3}\delta_{0,\nu_2}\delta_{0,\nu_3}z_0\,.
\label{eq:VTR11} \ea
Terms in which only two of the graviton fields are contracted
directly through a single contraction are grouped together into
the vertex
\ba V&\!\!\!\!\!\!^{TR,10}_{\mu_1 \nu_1,\mu_2\nu_2,\mu_3
\nu_3}&\!\!\!\!\!\!(Q_1,Q_2,Q_3)
\nonumber \\[2mm]
&=&
\frac{1}{4} \delta_{\mu_1,\sigma_2} \delta_{\nu_1,\nu_2}
 \delta_{\mu_2,\sigma_1} \delta_{\mu_3,\nu_3} Q_{1,\sigma_1} Q_{2,\sigma_2} z_0^3
+\frac{1}{2}
 \delta_{\mu_1,\sigma_1} \delta_{\nu_1,\nu_2} \delta_{\mu_2,\sigma_2} \delta_{\mu_3,\nu_3}
 Q_{1,\sigma_1} Q_{2,\sigma_2} z_0^3
\nonumber \\[2mm] &&
+\delta_{\mu_1,\sigma_1} \delta_{\nu_1,\nu_2} \delta_{\mu_2,\sigma_2} \delta_{\mu_3,\nu_3} Q_{2,\sigma_1}
 Q_{2,\sigma_2} z_0^3
+\delta_{\mu_1,\sigma_2} \delta_{\nu_1,\mu_2} \delta_{\nu_2,\sigma_1} \delta_{\mu_3,\nu_3} Q_{2,\sigma_1}
 Q_{3,\sigma_2} z_0^3
\nonumber \\[2mm] &&
+\delta_{\mu_1,\sigma_1} \delta_{\nu_1,\mu_2} \delta_{\nu_2,\sigma_2}
 \delta_{\mu_3,\nu_3} Q_{2,\sigma_1} Q_{3,\sigma_2} z_0^3
+\delta_{\mu_1,\sigma_1}
 \delta_{\nu_1,\mu_2} \delta_{\nu_2,\sigma_2} \delta_{\mu_3,\nu_3} Q_{3,\sigma_1} Q_{3,\sigma_2} z_0^3
\nonumber \\[2mm] &&
+2
 \delta_{\mu_1,\mu_2} \delta_{\nu_2,\sigma_1} \delta_{\mu_3,\nu_3} \delta_{0,\nu_1} Q_{2,\sigma_1} z_0^2
+\delta_{\mu_1,\sigma_1} \delta_{\nu_1,\mu_2} \delta_{\mu_3,\nu_3} \delta_{0,\nu_2}
 Q_{2,\sigma_1} z_0^2
\nonumber \\[2mm] &&
+4 \delta_{\nu_1,\nu_2} \delta_{\mu_2,\sigma_1}
\delta_{\mu_3,\nu_3} \delta_{0,\mu_1} Q_{3,\sigma_1} z_0^2
+\delta_{\mu_1,\mu_2} \delta_{\mu_3,\nu_3} \delta_{0,\nu_1}
\delta_{0,\nu_2} z_0\,. \ea
What remains are those terms of the three graviton vertex that
contain no direct contractions of two different graviton fields,
\ba V&\!\!\!\!\!\!^{TR,00}_{\mu_1 \nu_1,\mu_2\nu_2,\mu_3
\nu_3}&\!\!\!\!\!\!(Q_1,Q_2,Q_3)
\nonumber \\[2mm]
&=&
-\frac{1}{8} \delta_{\mu_1,\sigma_1} \delta_{\nu_1,\sigma_2} \delta_{\mu_2,\nu_2} \delta_{\mu_3,\nu_3}
 Q_{1,\sigma_1} Q_{1,\sigma_2} z_0^3
+\frac{1}{8} \delta_{\mu_1,\nu_1} \delta_{\mu_2,\nu_2}
 \delta_{\mu_3,\nu_3} \delta_{\sigma_1,\sigma_2} Q_{1,\sigma_1} Q_{1,\sigma_2} z_0^3
\nonumber \\[2mm] &&
-\frac{1}{2} \delta_{\mu_1,\sigma_1} \delta_{\nu_1,\sigma_2} \delta_{\mu_2,\nu_2} \delta_{\mu_3,\nu_3} Q_{1,\sigma_1} Q_{2,\sigma_2} z_0^3
+\frac{1}{8}
 \delta_{\mu_1,\nu_1} \delta_{\mu_2,\nu_2} \delta_{\mu_3,\nu_3} \delta_{\sigma_1,\sigma_2}
 Q_{1,\sigma_1} Q_{2,\sigma_2} z_0^3
\nonumber \\[2mm] &&
-\frac{1}{2} \delta_{\mu_1,\sigma_1} \delta_{\nu_1,\sigma_2}
 \delta_{\mu_2,\nu_2} \delta_{\mu_3,\nu_3} Q_{2,\sigma_1} Q_{2,\sigma_2} z_0^3
-\frac{1}{4} \delta_{\mu_1,\sigma_2} \delta_{\nu_1,\sigma_1}
 \delta_{\mu_2,\nu_2} \delta_{\mu_3,\nu_3} Q_{2,\sigma_1} Q_{3,\sigma_2} z_0^3
\nonumber \\[2mm] &&
+\frac{1}{2} \delta_{\mu_1,\nu_1}
 \delta_{\mu_2,\nu_2} \delta_{\mu_3,\nu_3} \delta_{0,\sigma_1} Q_{1,\sigma_1} z_0^2
-2 \delta_{\mu_1,\sigma_1} \delta_{\mu_2,\nu_2} \delta_{\mu_3,\nu_3} \delta_{0,\nu_1} Q_{2,\sigma_1} z_0^2
\nonumber \\[2mm] &&
-\frac1{2}z_0^2 \delta_{\mu_1,\nu_1} \delta_{\mu_2,\nu_2} \delta_{0,\nu_1} Q_{3,\mu_3}
+\frac{1}{6}
 \delta_{\mu_1,\nu_1} \delta_{\mu_2,\nu_2} \delta_{\mu_3,\nu_3} z_0
\nonumber \\[2mm] &&
-\frac{3}{4} \delta_{\mu_2,\nu_2}
 \delta_{\mu_3,\nu_3} \delta_{0,\mu_1} \delta_{0,\nu_1} z_0\,.
\ea
\begin{figure}
\begin{center}
{
\epsfysize4.0cm \epsfbox{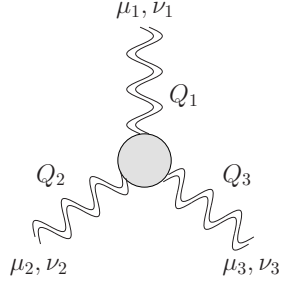}}
\end{center}
\caption{The three graviton vertex.}
\lab{fig:3gv}
\end{figure}

The symbols $Q_k$ denote five dimensional derivatives acting on
$k$-th external graviton propagators ($k$ runs from $1$ to $3$,
cf.\ fig.~\ref{fig:3gv}),
\be Q_{k,\mu} \equiv \partial_{{z_k}_\mu}\,. \ee
Before turning to the high energy limit, we still have to
symmetrize these expressions. This can be done in two steps. To
begin with, we symmetrize the two indices  $(\mu_k,\nu_k)$ for
each graviton (labelled by $k$). Then, in a second step, we also
symmetrize in the label $k$.

\subsection{The triple Regge limit}

So far we have worked in configuration space. The Fourier
transform is defined as before, and derivatives in configuration
space, as before, turn into external momenta, $\vec
p_1$,\ldots,$\vec p_6$. When computing the scattering amplitude in
the triple Regge limit one notices that the large energy
variables, $s_1$ and $s_2$, are constructed by contracting large
momenta contained in the stress-energy tensors via Kronecker
deltas from the graviton propagators and from the triple graviton
vertex. Since the graviton vertex involves at most two
contractions of external indices from two different gravitons, the
amplitude with the triple graviton vertex provides terms
proportional to $s_1^2$, $s_2^2$, or $s_1s_2$ plus lower order
contributions. In fact, the leading contribution from the triple
Regge limit comes form the terms (\ref{eq:VTR11}) and
(\ref{eq:VTR20}). While the former leads to terms which are
proportional to $s_1s_2$, the latter provides two types of terms
which are either proportional to $s_1^2$ or to $s_2^2$.

We compare this result with one expects from general
arguments~\cite{Brower:1974yv}. In the notation of Regge theory,
the kinematic limit which we referred to as the 'triple Regge
limit' is a mixed Regge-helicity limit. For this high energy limit
the Steinmann relations allow for four sets of non-vanishing
energy discontinuities. Following the arguments
in~\cite{Brower:1974yv} as well as eq.~(4.24) of the same paper,
one expects the six-point scattering amplitude to consist of four
terms. If we label the leading angular momentum singularities in
the three $t$ channels by $j$, $j_1$, and $j_2$, respectively, the
four terms have the following energy dependence
\begin{center}
\begin{tabular}{llclll}
  & & (i) &
 $(M^2)^{j- j_1 -j_2} s_1^{j_1} s_2^{j_2}$\,,
& and &  \\[2mm] (ii)& $s_2^{j}$\,,
 & (iii) &
 $s_1^{(j + j_1 -j_2)/2} s_2^{(j+ j_2 -j_1)/2}$\, , &
(iv) &
 $s_1^{j}$\,,
\\
\end{tabular}
\end{center}
The only term which contributes to the discontinuity in $M^2$ is
the first one: This is the six-point amplitude in QCD (or ${\cal
N}=4$ SYM) which we have described in the introduction. In the
weak coupling limit, the leading singularities in the angular
momentum plane are given by the BFKL Pomeron. Returning to
graviton exchange we have computed the complete (i.e.\ not
restricted ourselves to the $M^2$-dependent piece) six-point
correlator in the supergravity approximation. The leading
singularities are at $j=j_1=j_2=2$, and the three terms we have
found are in agreement with the energy dependence of (ii) - (iv).
The first term is absent, i.e.\ in the Witten diagram with
`elementary' graviton exchange, the triple graviton vertex is
found to vanish.

From the point of view of Feynman diagrams, this result can also
be understood as follows. In~\cite{Bartels:2009sc} it has been
demonstrated that the helicity structure of graviton exchange at
high energies can be viewed as the exchange of two spin one
bosons, each of them being in a circular polarized t-channel
helicity state. Correspondingly, in our high energy limit where
the graviton exchanges above and below the triple vertex can be
viewed as double-boson exchanges, the triple graviton vertex acts
like a product of two triple boson vertices. A simple look at the
triple gluon boson vertex of QCD shows that - in the triple Regge
limit - the six-point amplitude with three gluon exchange comes
with two terms: One of them is proportional to $s_1$ while the
other is proportional to $s_2$. Again, no term proportional to
$s_1 s_2 (M^2)^{-1}$ appears. Consequently, the product of two
such three gluon exchanges produces three terms, proportional to
$s_1^2$, $s_2^2$, and $s_1 s_2$.\footnote{It is interesting to
note that a nonzero triple vertex of reggeizing gluons in QCD has
been found~\cite{Hentschinski:2009ga}. After integration over
$M^2$ this vertex becomes zero, thus restoring signature
conservation.}

A similar result can also be found in flat supergravity
~\cite{Brandt:1992dk,Chen:1997vm}. In the zero slope limit the
triple graviton vertex decouples. A non-vanishing triple graviton
exchange is expected to appear only once the gravitons are
reggeized. This, however, requires a genuine string calculation
and thus goes beyond the scope of this paper.

\subsection{The coupling of two gravitons  and two $R-$bosons}

There is one more diagram we need to compute, namely the second
one depicted in fig. 4. In the high energy limit it will turn out
to contribute to the same order as the triple graviton exchange.
The analysis follows the same steps we have described at great
length in the first two subsections. Hence, we can be rather brief
now.  Copying our derivation of the triple graviton vertex, one
can calculate the vertex with two $R-$bosons and two gravitons,
i.e.\ the vertex that appears in the second diagram of
fig.~\ref{fig:grav3}. Making use of eqs.\
(\ref{eq:Rdef})-(\ref{eq:sqg}) we expand the kinetic term of
$R-$bosons
\be -\sqrt{g} F_{\mu  \nu }F^{\mu  \nu}= \sqrt{\bar
g}(F^{(0)}+F^{(1)}+F^{(2)})\,, \ee
where $F^{(0)} =   - z_0^4 F_{\mu  \nu }F_{\mu  \nu}$ and the
stress-energy tensor is defined by
\ba F^{(1)}\ = \  (2 z_0^2 F_{\rho  \mu }F_{\rho  \nu} -\frac1{2}
z_0^2 F_{\sigma  \rho }F_{\sigma  \rho} \delta_{\mu \nu}) z_0^4
h_{\mu \nu}\ = \  T_{\mu \nu} z_0^4 h_{\mu \nu} \,. \ea
The coupling of two gravitons and two $R-$bosons can be read from
\ba F^{(2)}&=&\left( \frac1{4}  F_{\rho \sigma}F_{\rho \sigma}
\delta_{\mu_1 \mu_2} \delta_{\nu_1 \nu_2}
-\frac1{8}  F_{\rho \sigma}F_{\rho \sigma}
\delta_{\mu_1 \nu_1} \delta_{\mu_2 \nu_2}
+\frac1{2}  F_{\rho \mu_2}F_{\rho \nu_2}
\delta_{\mu_1 \nu_1}
\right. \nonumber \\[2mm] &&\left.
+\frac1{2}  F_{\rho \mu_1}F_{\rho \nu_1}
\delta_{\mu_2 \nu_2}
-  F_{\mu_1 \nu_2}F_{\nu_1 \mu_2}
-2  F_{\mu_1 \mu_2}F_{\nu_1 \nu_2}
\right)
z_0^8 h_{\mu_1 \nu_1}h_{\mu_2 \nu_2}\,.
\lab{eq:cGGFF}
\ea
In the high energy limit, the diagram under consideration can only
give subleading contribution which are proportional to $s_1^2$,
$s_2^2$, or $s_1 s_2$. In fact, as we have argued previously,
powers of $s_1$ and $s_2$ appear if and only if momenta
(derivatives) from the field strength tensors $F_{\mu \nu}$ are
contracted by the Kronecker deltas coming with the graviton
propagators. In the coupling (\ref{eq:cGGFF}) of two gravitons and
two $R-$bosons, each term involves only two field strength
tensors. Since each field strength tensor contains only one
momentum that is contracted with the graviton by using eqs.\
(\ref{eq:pG1})-(\ref{eq:pG0}), contributions proportional to
$s_1^2 s_2^2$ are impossible to obtain. The first two terms of the
vertex lead to traces over the graviton propagator and hence they
furnish constant contributions to high energy scattering. The
remaining terms behave as $s_i s_j$, at most. Hence, at high
energies, the six-point correlator of $R$-currents is dominated by
the two diagrams in fig. \ref{fig:gravboson}. The two diagrams in
fig. \ref{fig:grav3} are subleading.

\section{Summary}
In this paper we have investigated the correlation function of six
$R$-currents at high energies and in the strong coupling limit.
Interest in such six-point functions comes from the observation
that graviton exchanges at high energies need to be unitarized. As
a first step, we need to compute the coupling of two gravitons to
the $R$-current. Such a coupling appears as a part of the
six-point function. We have two classes of Witten diagrams, one
containing the two graviton exchanges depicted in
fig.~\ref{fig:gravboson}, the other one containing the three
graviton exchange in fig.~\ref{fig:grav3}. The latter one
represents the triple Regge limit. These Witten diagrams have
their analogues on the weak coupling side, i.e.\ in the high
energy behavior of $R$-current correlators in ${\cal N}=4$ SYM:
The diagrams in fig.~\ref{fig:gravboson} correspond to the
exchange of two BFKL Pomerons on the weak coupling side, see
fig.~\ref{fig:triplereggeSYM}, left figure. On the other hand, the
triple graviton diagram in fig.~\ref{fig:grav3} has its weak
coupling counterpart in the triple Pomeron diagram on the right
hand side of fig.~\ref{fig:triplereggeSYM}. It is remarkable that
the existence of the former contribution is a consequence of the
supersymmtric structure of ${\cal N}=4$ SYM, and it does not hold
for (nonsupersymmetric) QCD. The study of the present paper can be
viewed as the strong coupling analogue of an earlier
paper~\cite{Bartels:2009ms}.

Beginning with the two graviton exchange, the correlation function
has the same structure as on the weak coupling side, a convolution
of impact factors and exchange propagators. The integration is
over the position of the impact factors in the direction of the
fifth coordinate. One of our main results is the new impact factor
which describes the coupling of two gravitons to the upper
$R$-boson. Similar to its weak coupling counterpart (which
consists of a closed loop of spinors and scalars in the adjoint
representation of the color group), it has a cut in the mass
variable $M^2$, is maximal for small $M^2$ and, for large $M^2$,
falls off as $M^{-4}$.

In the second part we have considered the three graviton diagram.
We derived an expression for the triple graviton vertex, and found
that the coupling of three elementary gravitons vanishes in the
triple Regge limit. In agreement with the Steinmann relations, we
obtained three terms which grow as $s_1^2$, $s_2^2$, and $s_1s_2$,
respectively. We expect that the triple graviton vertex will be
nonzero once the attached gravitons reggeize. This, however,
requires genuine string scattering amplitudes and thus goes well
beyond the analysis of Witten diagrams. Note that the triple
vertex of the BFKL Pomeron in weakly coupled QCD possesses a
non-trivial inner structure. This is linked to the fact that the
BFKL Pomeron is a composite object. Hence, it is tempting to
expect some kind of reggeization for the dual graviton so as to
match its triple vertex with that of the Pomeron.

As we have said at the beginning, our present study was mainly
motivated by the interest in two-graviton exchange. As a first
step, we have investigated the coupling of two gravitons to the
$R$-current. The existence of the direct coupling hints at the
importance of eikonalization. Nevertheless, the triple graviton
diagram also needs further investigation.

Our study of higher order $R$-current correlators should be seen
also within another context. One of the most important ingredients
in the analysis of gauge/string dualities is the remarkable
appearance of integrability. For multi-color QCD is was shown many
years ago, see  ~\cite{Lipatov:1993yb,Lipatov:1994xy,Faddeev:1994zg}, that the
BKP Hamiltonian, i.e.\ the operator that encodes the rapidity
evolution of $n$-gluon $t$ channel states, corresponds to a closed
spin chain and is integrable. Such BKP states enter the high
energy limit of scattering amplitudes with more than eight
external legs. Our study of the six-point amplitude therefore also
serves as a preparation for pursuing further studies in this
direction.

\section*{Acknowledgments}
We are grateful for discussions with A.~H.~Mueller, G.~P.~Vacca and L.~Motyka.
This work was supported by the grant of
SFB 676, Particles, Strings and the Early Universe:
``the Structure of Matter and Space-Time''.

\appendix

\section{Integrals for the forward case}
\lab{sc:int}
To calculate the forward case as well as
the OPE limit we have found the following integrals
\be \gamma \int_v^\infty d v_A
 v_A
 K_{0}(v_A )
K_{0}(v_A \gamma) \ = \ - \gamma \frac{v \left(K_0(\gamma v)
K_1(v)-\gamma K_0(v) K_1(\gamma v)\right)}{(\gamma-1) (\gamma+1)}
\ = \ \frac{ \gamma \log (\gamma)}{\gamma^2-1}+O\left(v^2\right)
\ee and\ba \gamma \int_v^\infty d v_A
 v_A
 K_{1}(v_A )
K_{1}(v_A \gamma)
&=& \gamma
\frac{v \left(\gamma K_0(\gamma v) K_1(v)-K_0(v) K_1(\gamma v)\right)}{
(\gamma+1)(\gamma-1)}
\nonumber \\[2mm]
&=& \log (v^{-1}) + (\log (2)-\gamma_E) +\frac{\gamma^2 \log
\left(\gamma\right) }{\gamma^2-1} +O\left(v^2\right) \ea as well
as \ba \int_0^v &\!\!\!\!\!\!d v_A &\!\!\!\!\!\! v_A^5 K_{0}(v_A )
K_{0}(v_A \gamma) \ = \ \left(-\frac{4 \left(\gamma ^2+1\right)
v^4}{\left(\gamma
 ^2-1\right)^2}-\frac{32 \left(\gamma ^4+4 \gamma ^2+1\right)
 v^2}{\left(\gamma ^2-1\right)^4}\right)
K_0(v) K_0(\gamma v)
\nonumber \\[2mm] &&
+\left(\frac{v^5}{\gamma ^2-1}+\frac{16 \left(2 \gamma ^2+1\right)
 v^3}{\left(\gamma ^2-1\right)^3}+\frac{64 \left(\gamma ^4+4 \gamma
 ^2+1\right) v}{\left(\gamma ^2-1\right)^5}\right) K_1(v) K_0(\gamma
 v)
\nonumber \\[2mm] &&
+\left(\frac{\gamma v^5}{1-\gamma ^2}-\frac{16 \gamma \left(\gamma
 ^2+2\right) v^3}{\left(\gamma ^2-1\right)^3}-\frac{64 \left(\gamma ^5+4
 \gamma ^3+\gamma \right) v}{\left(\gamma ^2-1\right)^5}\right) K_0(v)
 K_1(\gamma v)
\nonumber \\[2mm] &&
+\left(\frac{8 \gamma v^4}{\left(\gamma
 ^2-1\right)^2}+\frac{96 \left(\gamma ^3+\gamma \right)
 v^2}{\left(\gamma ^2-1\right)^4}\right) K_1(v) K_1(\gamma v)
\nonumber \\[2mm] && +\frac{32\left(-3 \gamma ^4+2 \left(\gamma ^4+4
\gamma ^2+1\right) \log (\gamma )+3\right)}{\left(\gamma
^2-1\right)^5} \ea and \ba \int_0^v &\!\!\!\!\!\!d v_A
&\!\!\!\!\!\! v_A^5 K_{1}(v_A ) K_{1}(v_A \gamma) \ = \
\left(\frac{8 \gamma v^4}{\left(\gamma ^2-1\right)^2}+\frac{96
 \left(\gamma ^3+\gamma \right) v^2}{\left(\gamma ^2-1\right)^4}\right)
 K_0(v) K_0(\gamma v)
\nonumber \\[2mm] &&
+\left(\frac{\gamma v^5}{1-\gamma ^2}-\frac{8
 \gamma \left(\gamma ^2+5\right) v^3}{\left(\gamma
 ^2-1\right)^3}-\frac{192 \left(\gamma ^3+\gamma \right) v}{\left(\gamma
 ^2-1\right)^5}\right) K_1(v) K_0(\gamma v)
\nonumber \\[2mm] &&
+\left(\frac{v^5}{\gamma
 ^2-1}+\frac{8 \left(5 \gamma ^2+1\right) v^3}{\left(\gamma
 ^2-1\right)^3}+\frac{192 \left(\gamma ^4+\gamma ^2\right)
 v}{\left(\gamma ^2-1\right)^5}\right) K_0(v) K_1(\gamma
 v)
\nonumber \\[2mm] &&
+\left(-\frac{4 \left(\gamma ^2+1\right) v^4}{\left(\gamma
 ^2-1\right)^2}-\frac{16 \left(\gamma ^4+10 \gamma ^2+1\right)
 v^2}{\left(\gamma ^2-1\right)^4}\right) K_1(v) K_1(\gamma v)
\nonumber \\[2mm] &&
+\frac{16
 \left(\gamma ^6+9 \gamma ^4-9 \gamma ^2-12 \left(\gamma ^4+\gamma
 ^2\right) \log (\gamma )-1\right)}{\gamma \left(\gamma ^2-1\right)^5}
\ea
The above results can be also used to perform integrals
from~\cite{Bartels:2009sc}.

\section{Integrals appearing in the DIS limit}
\lab{sc:disI}

In this appendix we present further details of the six-point
amplitude, restricting ourselves to the limit of deep inelastic
scattering. We will be slightly more general than in section 3.4,
by allowing the external virtualities to be less restricted. In
particular, we allow $|\vec p_1|, |\vec p_4|\gg |\vec p_2|, |\vec
p_3|, |\vec p_5|, |\vec p_6|$, without the constraints   $|\vec
p_1| = |\vec p_4|$ etc., and we define
\be \alpha\ = \ |\vec p_1|/M \,, \quad \beta\ = \ |\vec p_2|/|\vec
p_1|\,, \quad \rho\ = \ |\vec p_4|/|\vec p_1|\,,
\quad \rho_1\ = \ |\vec p_5|/|\vec p_2|\,, \quad \rho_2\ = \ |\vec
p_6|/|\vec p_3|\,. \ee
As a result, our integrals depend also upon the variables $\rho$,
$\rho_1$, $\rho_2$. Thus, the exchange defined by planar diagram
reads as
\ba {\cal A}^{\rm 2G,planar}_{\lambda_A \lambda_{B1}\lambda_{B2}}
&\approx& - M^{-2} \left(\frac{s_1}{|\vec p_1| |\vec
p_4|}\right)^2 \left(\frac{s_2}{|\vec p_1| |\vec p_4|}\right)^2
I_{\lambda}(-\alpha^2,\rho) L_{\lambda_{B1}}(\beta,\rho_1)
L_{\lambda_{B2}}(\beta,\rho_2)\,, \,
\lab{eq:fAplan} \ea where the integrations over {\em lower}
vertices give \be L_{\lambda_{B}}(\beta,\rho)\ = \
\log^{m(\lambda_{B})}(\beta^{-2}) \left(\frac{\rho \log
(\rho^2)}{\rho^2-1}\right)^{1-m(\lambda_{B})} \ee
while contribution coming from the integral over {\em upper}
vertices, $I_\lambda(-\alpha^2,\rho)$, is defined by
\be - \alpha^2 \rho I_\lambda(-\alpha^2,\rho)\ = \
p^{(0)}_\lambda + p^{(1)}_\lambda \log(-\alpha^2) +
p^{(2)}_\lambda \log(\rho)\,. \ee
%
For the transverse polarization we found that
\ba p^{(0)}_T&=&\frac{96 \alpha ^2 \rho ^4 }{\left(\alpha
^2+1\right)^4 \left(\rho ^2-1\right)^8 \left(\alpha ^2 \rho
^2+1\right)^4} \nonumber \\[2mm] && \left( \rho ^5 \left(\rho
^2+1\right) \left(\rho ^{12}-9 \rho ^{10}+17 \rho ^8-858 \rho
^6+17 \rho ^4-9 \rho ^2+1\right) \alpha ^{14} \right. \nonumber \\[2mm]
&&\left. -2 \rho ^3 \left(5 \rho ^{16}-39 \rho ^{14}+172 \rho
^{12}+1333 \rho ^{10}+2938 \rho ^8+1333 \rho ^6+172 \rho ^4-39
\rho ^2+5\right) \alpha
 ^{12}
\right. \nonumber \\[2mm] &&\left.
+\rho \left(\rho ^2+1\right) \left(\rho ^{16}\!+11 \rho
^{14}\!-261 \rho ^{12}\!-4081 \rho ^{10}\!-8980 \rho ^8\!-4081
\rho ^6\!-261 \rho ^4\!+11 \rho
 ^2+1\right) \alpha ^{10}
\right. \nonumber \\[2mm] &&\left.
-2 \left(\rho ^{17}+42 \rho ^{15}+1609 \rho ^{13}+7020 \rho ^{11}+12056 \rho ^9+7020 \rho ^7+1609 \rho ^5+42 \rho ^3+\rho
 \right) \alpha ^8
\right. \nonumber \\[2mm] &&\left.
-\rho \left(\rho ^2+1\right) \left(18 \rho ^{12}+883 \rho ^{10}+6856 \rho ^8+13886 \rho ^6+6856 \rho ^4+883 \rho ^2+18\right) \alpha
 ^6
\right. \nonumber \\[2mm] &&\left.
-4 \rho \left(8 \rho ^{12}+437 \rho ^{10}+2125 \rho ^8+3680 \rho ^6+2125 \rho ^4+437 \rho ^2+8\right) \alpha ^4
\right. \nonumber \\[2mm] &&\left.
-\rho \left(\rho ^2+1\right)
 \left(23 \rho ^8+1298 \rho ^6+3238 \rho ^4+1298 \rho ^2+23\right) \alpha ^2
\right. \nonumber \\[2mm] &&\left.
-2 \rho \left(3 \rho ^8+178 \rho ^6+478 \rho ^4+178 \rho
 ^2+3\right)\right)
\ea
\ba
p^{(1)}_T&=&\frac{576 \alpha ^{14} \left(\alpha ^2-1\right) \rho ^7 \left(\alpha ^2 \rho ^2-1\right)}{\left(\alpha ^2+1\right)^5 \left(\alpha ^2 \rho ^2+1\right)^5}
\ea
\ba
p^{(2)}_T&=&\frac{1152 \alpha ^2 \rho ^7
}{\left(\rho ^2-1\right)^9 \left(\alpha ^2 \rho
 ^2+1\right)^5}
\nonumber \\[2mm] &&
\left(
10 \alpha ^8 \left(5 \rho ^4+18 \rho ^2+5\right) \rho ^{10}
+\alpha ^6 \left(145 \rho ^{10}+669 \rho ^8+334 \rho ^6-36
 \rho ^4+9 \rho ^2-1\right) \rho ^2
\right. \nonumber \\[2mm] &&\left.
+20 \left(\rho ^6+6 \rho ^4+6 \rho ^2+1\right)+\alpha ^2 \left(94 \rho ^8+534 \rho ^6+464 \rho ^4+34 \rho
 ^2-6\right)
\right. \nonumber \\[2mm] &&\left.
+\alpha ^4 \left(171 \rho ^{10}+897 \rho ^8+632 \rho ^6-18 \rho ^4-3 \rho ^2+1\right)\right)
\ea
while for the longitudinal polarization
\ba p^{(0)}_L&=&-\frac{192 \alpha ^2 \rho ^6 }{\left(\alpha
^2+1\right)^4 \left(\rho ^2-1\right)^8 \left(\alpha ^2 \rho
^2+1\right)^4} \nonumber \\[2mm] && \left( \rho \left(\rho ^{17}-8 \rho
^{15}+28 \rho ^{13}-186 \rho ^{11}-510 \rho ^9-186 \rho ^7+28 \rho
^5-8 \rho ^3+\rho
 \right) \alpha ^{14}
\right. \nonumber \\[2mm] &&\left.
-\left(\rho ^2+1\right) \left(\rho ^{16}-4 \rho ^{14}-8 \rho ^{12}+612 \rho ^{10}+1738 \rho ^8+612 \rho ^6-8 \rho ^4-4 \rho
 ^2+1\right) \alpha ^{12}
\right. \nonumber \\[2mm] &&\left.
+\left(5 \rho ^{16}-34 \rho ^{14}-688 \rho ^{12}-4552 \rho ^{10}-7102 \rho ^8-4552 \rho ^6-688 \rho ^4-34 \rho ^2+5\right)
 \alpha ^{10}
\right. \nonumber \\[2mm] &&\left.
-2 \left(\rho ^2+1\right) \left(3 \rho ^{12}+236 \rho ^{10}+1704 \rho ^8+3464 \rho ^6+1704 \rho ^4+236 \rho ^2+3\right) \alpha ^8
\right. \nonumber \\[2mm] &&\left.
-2
 \left(62 \rho ^{12}+791 \rho ^{10}+3653 \rho ^8+5688 \rho ^6+3653 \rho ^4+791 \rho ^2+62\right) \alpha ^6
\right. \nonumber \\[2mm] &&\left.
-6 \left(\rho ^2+1\right) \left(41 \rho ^8+336
 \rho ^6+716 \rho ^4+336 \rho ^2+41\right) \alpha ^4
\right. \nonumber \\[2mm] &&\left.
-2 \left(93 \rho ^8+743 \rho ^6+1268 \rho ^4+743 \rho ^2+93\right) \alpha ^2
\right. \nonumber \\[2mm] &&\left.
-10 \left(\rho
 ^2+1\right) \left(5 \rho ^4+32 \rho ^2+5\right)\right)
\ea
\ba
p^{(1)}_L&=&\frac{64 \alpha ^{12}
\left(\alpha ^4-4 \alpha ^2+1\right) \rho ^6 \left(\alpha ^4 \rho ^4-4 \alpha ^2 \rho ^2+1\right)}{\left(\alpha ^2+1\right)^5
 \left(\alpha ^2 \rho ^2+1\right)^5}
\ea
\ba
p^{(2)}_L&=&-\frac{128 \alpha ^2 \rho ^6
}{\left(\rho ^2-1\right)^9 \left(\alpha ^2 \rho ^2+1\right)^5}
\nonumber \\[2mm] &&
\left(\rho ^4 \left(100 \rho ^{12}+1125 \rho ^{10}+1251 \rho ^8+16 \rho ^6+36 \rho ^4-9 \rho ^2+1\right) \alpha ^8
\right. \nonumber \\[2mm] &&\left.
+\rho ^2
 \left(275 \rho ^{12}+3681 \rho ^{10}+5652 \rho ^8+512 \rho ^6-63 \rho ^4+27 \rho ^2-4\right) \alpha ^6
\right. \nonumber \\[2mm] &&\left.
+\left(316 \rho ^{12}+4617 \rho ^{10}+8523 \rho
 ^8+1888 \rho ^6-252 \rho ^4+27 \rho ^2+1\right) \alpha ^4
\right. \nonumber \\[2mm] &&\left.
+9 \left(19 \rho ^{10}+293 \rho ^8+608 \rho ^6+208 \rho ^4-7 \rho ^2-1\right) \alpha ^2
\right. \nonumber \\[2mm] &&\left.
+36
 \left(\left(\rho ^2+2\right) \left(\rho ^4+14 \rho ^2+8\right) \rho ^2+1\right)\right)
\ea
One can notice that the poles in $\alpha^2$-plane are spurious,
i.e.\ all poles of $p^{(k)}_\lambda(\alpha^2,\rho)$ cancel each
other in the sum.

The contribution related to the crossed diagram is defined by the
formula with $-M^2 \to \tilde M^2\equiv M^2+t-t_1-t_2+|\vec
p_1|^2+|\vec p_4|^2$, namely ${\cal A}^{\rm 2G,crossed}_{\lambda_A
\lambda_{B1}\lambda_{B2}}$ is analytic continuation of ${\cal
A}^{\rm 2G,planar}_{\lambda_A \lambda_{B1}\lambda_{B2}}$ in $M^2$
plane. In the large $M^2$ limit the leading terms of ${\cal
A}^{\rm 2G,crossed}_{\lambda_A \lambda_{B1}\lambda_{B2}}$, which
is of $\tilde M^{-2} \approx M^{-2}$ order, cancels with the
leading term of ${\cal A}^{\rm 2G,planar}_{\lambda_A
\lambda_{B1}\lambda_{B2}}$. This means that the sum is of $M^{-4}$
order
\ba {\cal A}^{\rm 2G,planar}_{\lambda_A \lambda_{B1}\lambda_{B2}}
+ {\cal A}^{\rm 2G,crossed}_{\lambda_A \lambda_{B1}\lambda_{B2}}
&=&- \frac{|\vec p_1| |\vec p_4| }{M^4} \left(\frac{s_1}{|\vec
p_1| |\vec p_4|}\right)^2 \left(\frac{s_2}{|\vec p_1| |\vec
p_4|}\right)^2 \hat I_{\lambda}(\alpha^2,\rho)
L_{\lambda_{B1}}(\beta,\rho_1) L_{\lambda_{B2}}(\beta,\rho_2)\,,
\nonumber \\[2mm]
\lab{eq:fAtot} \ea
where $t_1=t_2=t=0$. The function describing the sum of the planar
and crossed {\em upper} impact factor reads as
\be \hat I_{\lambda}(\alpha^2,\rho) \ = \
 \alpha^{-2} \rho^{-1}
(I_{\lambda}(- \alpha^2,\rho) -
I_{\lambda}((\alpha^{-2}+1+\rho^2)^{-1},\rho))\,, \ee
In the large $M^2$ limit, its value is defined by
\be \hat I_{\lambda}(\alpha^2=0,\rho) \ = \ \rho^4 \int_0^{\infty}
dr r^7 K_{\epsilon_\lambda}(r) K_{\epsilon_\lambda}(r \rho) \,,
\ee
and it is plotted in fig.~\ref{fig:hI} as a function of the ratio
of {\em upper} virtualities, i.e.\ $\rho$. The function reminds
the Gaussian profile with maximum at $|\vec p_1|=|\vec p_4|$.
\begin{figure}
\begin{center}
{
\psfrag{'mtnrrt.dat'}{$\lambda=T$}
\psfrag{'mtnrrl.dat'}{$\lambda=L$}
\psfrag{r}{$\frac{1-\rho}{1+\rho}$}
\epsfysize7.0cm \epsfbox{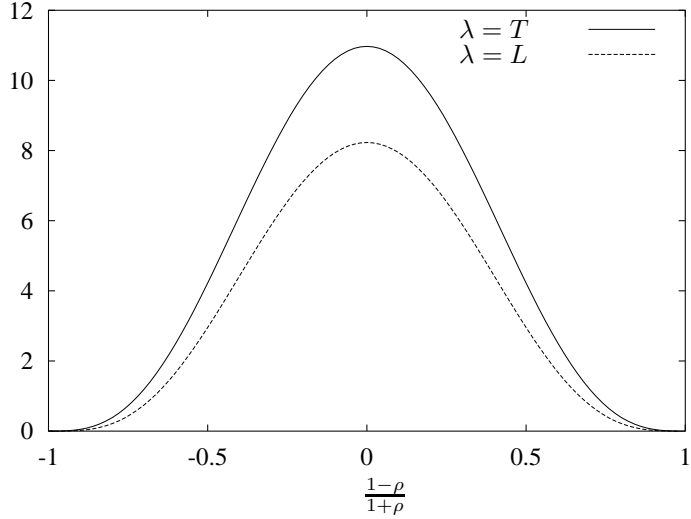}
}
\end{center}
\caption{Functions
$\hat I_{\lambda}(\alpha^2=0,\rho)=(\rho+\rho^{-1}) I^{(0)}_{\lambda}(\rho)
-2 I^{(1)}_{\lambda}(\rho)$ plotted as a function of
$\frac{1-\rho}{1+\rho}=\frac{|\vec p_1|-|\vec p_4|}{|\vec p_1|+|\vec p_4|}$}
\lab{fig:hI}
\end{figure}

Making use of eq.\ (\ref{eq:KIep}) one can find that the imaginary
part of $R$-boson propagator
\be \Im \tilde {\cal
K}_{\epsilon_\lambda}(z_M/\alpha,y_M/\alpha;\mp i) \ = \ \pm
\frac{\pi}{2} J_{\epsilon_\lambda} (z_M/\alpha)
J_{\epsilon_\lambda} (y_M/\alpha)\,. \ee
This allows to calculate simply the imaginary part of the
amplitude (\ref{eq:fAplan}) related to discontinuity along
$M^2>0$, namely
\ba {\rm disc}_{M^2} I_{T}(-\alpha^2,\rho)&=&- \frac{576 \alpha
^{12} \left(\alpha ^2-1\right) \pi \rho
 ^6 \left(\alpha ^2 \rho ^2-1\right)}{\left(\alpha
 ^2+1\right)^5 \left(\alpha ^2 \rho ^2+1\right)^5}
\lab{eq:fiIT}
\ea
and
\ba {\rm disc}_{M^2} I_{L}(-\alpha^2,\rho)&=&- \frac{64 \alpha
^{10} \left(\alpha ^4-4 \alpha
 ^2+1\right) \pi \rho ^5 \left(\alpha ^4 \rho ^4-4
 \alpha ^2 \rho ^2+1\right)}{\left(\alpha ^2+1\right)^5
 \left(\alpha ^2 \rho ^2+1\right)^5}
\lab{eq:fiIL}
\ea
The roots, which are related to the change of the amplitude phase,
appear at
\ba
 |\vec p_1|/M, |\vec p_4|/M &= 1 &
\mbox{for the transverse part,}
\\[2mm] \nonumber
 |\vec p_1|/M, |\vec p_4|/M & = \frac{1}{\sqrt{2}}(\sqrt{3}\pm 1) &
\mbox{for the longitudinal part.}
\\[2mm] \nonumber
\ea
Also, similarly to the $\rho=1$ case we can observe the symmetry of
\be
\rho^{-2} \alpha^{-4} \Im I_{\lambda}(-\alpha^2,\rho)
\quad
\mbox{
under
}
\quad
\rho \alpha^2 \leftrightarrow (\rho \alpha^2 )^{-1}\,,
\ee
where $\rho \alpha^2 \equiv \frac{|\vec p_1| |\vec p_4|}{ M^2}$.
Thus, the discontinuity multiplied by $(\rho\alpha^2)^{-3}$ is
invariant under the inversion in the $M^2/(|\vec p_1||\vec p_4|)$
variable.


\section{The saddle point method for large $M^2$ expansion}
\lab{sc:alt}

In this appendix we calculate the real part of the integral
\ba I_{\lambda}(\alpha^2,\rho) &=& \ft1{32} \alpha^{-2} \rho^5
\int_0^{\infty} d z_M \int_0^{\infty} d y_M z_M^5 y_M^5
K_{\epsilon_\lambda}(z_M) K_{\epsilon_\lambda}(y_M \rho ) \tilde
{\cal K}_{\epsilon_\lambda}(z_M/\alpha,y_M/\alpha;1 )\,,
\lab{eq:fIpra} \ea
from eq.\ (\ref{eq:Aplan}) in large $M^2$ limit making use
expression for the propagator $\tilde {\cal
K}_{\epsilon_\lambda}(z_M,y_M;1 )$ defined by eq.\ (\ref{eq:tKa}).
Let us change variables $z_M=r \sin(\phi)$ and $y_M=r \cos(\phi)$
and $|J|=r$. Analyzing eq.\ (\ref{eq:fIpra}) one can find that in
its first two order expansion in small $\alpha$ the leading
contribution comes from the region where $k \sim \alpha^{-2}$.
Thus we can apply the saddle point method with the large $k$
parameter, i.e.\
\ba I_{\lambda}(\alpha^2,\rho) &=& \int_0^{\infty} dr
\sum_{k=0}^{\infty} \alpha^{-\epsilon_\lambda-2 k-2}
\frac{4^{-k-3} (2-\epsilon_\lambda+2 (1-\epsilon_\lambda) k) }{
\Gamma (k+1) \Gamma (k+2)} r^{11+2k+\epsilon_\lambda} \rho ^5
\nonumber \\[2mm] && \hspace*{-10mm} \int_0^{\pi/2} d \phi K_{\epsilon_\lambda+2
k}\left(\frac{r}{\alpha }\right) K_{\epsilon_\lambda}(r \rho \cos
(\phi )) K_{\epsilon_\lambda}(r \sin (\phi )) \cos
^{5+{\epsilon_\lambda}}(\phi ) \sin ^{5+{\epsilon_\lambda}}(\phi )
(\sin (\phi ) \cos (\phi))^{2 k}
\nonumber \\[2mm]
&& \hspace*{-20mm} =\, \int_0^{\infty} dr \sum_{k=0}^{\infty}
\alpha^{-{\epsilon_\lambda}-2 k-2} \frac{4^{-k-3}
(2-{\epsilon_\lambda}+2 (1-{\epsilon_\lambda}) k) }{ \Gamma (k+1)
\Gamma (k+2)} r^{11+2k+{\epsilon_\lambda}} \rho ^5
K_{{\epsilon_\lambda}+2 k}\left(\frac{r}{\alpha }\right) h(r,k)\,,
\ea
where
\ba h(r,k) &=& \int_0^{\pi/2}  d \phi g_r(\phi) \e^{ k f(\phi)}
\, =\,  \int_0^{\pi/2}  d \phi \left( g_r(\phi_0)+
g_r'(\phi_0)(\phi-\phi_0)+ \ft{1}{2} g_r''(\phi_0)(\phi-\phi_0)^2
+\ldots \right)
\nonumber \\[2mm]
&& \hspace*{-5mm} \e^{ k f(\phi_0)+ k f'(\phi_0)(\phi-\phi_0)
+\ft{1}{2!} k f''(\phi_0)(\phi-\phi_0)^2 +\ft{1}{3!} k
f'''(\phi_0)(\phi-\phi_0)^3 +\ft{1}{4!} k
f^{(iv)}(\phi_0)(\phi-\phi_0)^4 +\ldots }\,,  \ea
with
\ba g_r(\phi) &=& K_{\epsilon_\lambda}(r \rho \cos (\phi ))
K_{\epsilon_\lambda}(r \sin (\phi )) \cos
^{5+{\epsilon_\lambda}}(\phi ) \sin ^{5+{\epsilon_\lambda}}(\phi
)\,, \ea
and
\ba f(\phi)&=&2\log(\sin (\phi ) \cos (\phi)) \, =\, 2 \log
\left(\cos (\phi_0) \sin(\phi_0)\right) +4 \cot (2 \phi_0) (\phi
-\phi_0)\nonumber \\[2mm] & & \hspace*{15mm} -4 \csc ^2(2 \phi_0) (\phi -\phi_0)^2
 +\frac{16}{3} \cot (2 \phi_0) \csc^2(2 \phi_0) (\phi -\phi_0)^3
\nonumber \\[2mm] & & \hspace*{15mm} -\frac{8}{3}\left((\cos (4 \phi_0)+2) \csc ^4(2
\phi_0)\right) (\phi -\phi_0)^4  +O\left((\phi
-\phi_0)^5\right)\,. \ea
Since we are going to calculate the first two orders we have to
expand
 $f(\phi)$ to fourth order and
 $g(\phi)$ to second order.
The saddle point corresponds to $z_0 = y_0$, i.e.\ $\phi_0=\pi/4$.
It is defined by $ \cot (2 \phi_0)=0$, so that
$f'(\phi_0)=f'''(\phi_0)=0$.
To integrate out $\phi$ we use
\ba \int_0^{\pi/2}  d \phi \e^{- 4 k \phi^2 -\ft{8}{3} k \phi^4}
&=&\frac{\sqrt{3}}{2 \sqrt{2}} \e^{\frac{3k}{4}}
K_{1/4}\left(\frac{3k}{4}\right)
\nonumber \\[2mm]
&=&
\frac{1}{2} \sqrt{\pi }
k^{-\frac{1}{2}}
-\frac{1}{16} \sqrt{\pi }
k^{-\frac{3}{2}}
+\frac{35}{768} \sqrt{\pi }
k^{-\frac{5}{2}}
+O\left(
k^{-\frac{7}{2}}
\right)\,,
\ea
and
\ba
\int_0^{\pi/2}  d \phi \phi^2 \e^{- 4 k \phi^2 -\ft{8}{3} k \phi^4}
&=&
\frac{3}{16} \sqrt{\frac{3}{2}} e^{3 k/4}
 \left(K_{\frac{3}{4}}\left(\frac{3
 k}{4}\right)-K_{\frac{1}{4}}\left(\frac{3
 k}{4}\right)\right)
\nonumber \\[2mm]
&=&
\frac{1}{16} \sqrt{\pi }
k^{-\frac{3}{2}}
-\frac{5}{128} \sqrt{\pi }
k^{-\frac{5}{2}}
+\frac{105 }{2048} \sqrt{\pi }
k^{-\frac{7}{2}}
+O\left(k^{-\frac{9}{2}}\right)\,,
\ea
which results
\ba
I_{\lambda}(\alpha^2,\rho)
&=&-
\int_0^{\infty} dr
\sum_{k=0}^{\infty}
\frac{2^{-\epsilon_\lambda -4 k-17}
}{\alpha ^2 k^{5/2} \Gamma (k) \Gamma (k+2)}
 (\epsilon_\lambda +2 (\epsilon_\lambda -1) k-2)
\sqrt{\pi } r^{11}
K_{\epsilon_\lambda +2 k}\left(\frac{r}{\alpha }\right)
\nonumber \\[2mm] &&
\left(\frac{r}{\alpha
 }\right)^{\epsilon_\lambda +2 k} \rho ^2 \left(2 r
 K_{\epsilon_\lambda +1}\left(\frac{r}{\sqrt{2}}\right)
 \left(\sqrt{2} (\epsilon_\lambda +1) K_{\epsilon_\lambda
 }\left(\frac{r \rho }{\sqrt{2}}\right)-r \rho
 K_{\epsilon_\lambda +1}\left(\frac{r \rho
 }{\sqrt{2}}\right)\right)
\right.
\nonumber \\[2mm] &&
\left.
+K_{\epsilon_\lambda
 }\left(\frac{r}{\sqrt{2}}\right)
 \left(\left(\left(\rho ^2+1\right) r^2-16 \epsilon_\lambda +32 k-44\right)
 K_{\epsilon_\lambda }\left(\frac{r \rho }{\sqrt{2}}\right)
\right.
\right.
\nonumber \\[2mm] &&
\left.
\left.
 +2 \sqrt{2} (\epsilon_\lambda +1) r \rho
 K_{\epsilon_\lambda +1}\left(\frac{r \rho }{\sqrt{2}}\right)
\right)\right)
+\ldots\,.
\ea

Since the dominant contribution for small $\alpha$ is defined in
the region where
\ba k\ = \  \kappa /\alpha^2 \quad \mbox{with} \quad \kappa\ = \
\mbox{fixed}\,, \ea
one can exchange sum over $k$ by integral over $\kappa$. We
substitute the large $k$ expansion of Bessel functions, i.e.\
\be K_{2k+\epsilon_\lambda}(r/\alpha) \approx \frac{1}{2}
\left(\frac{r}{2 \alpha }\right)^{-\epsilon_\lambda -2 k}
\sum_{j=0}^{J} \frac{ \Gamma (\epsilon_\lambda +2 k-j)}{\Gamma
(j+1)} \left(-\frac{r^2}{4 \alpha ^2}\right)^j\,, \ee
and making use of
\ba \frac{(2 k)^{j-{\epsilon_\lambda}}}{2^{-2 k}} \frac{\Gamma
({\epsilon_\lambda}-j+2 k)}{\Gamma (k) \Gamma (k+2)} &\approx&
\frac{\alpha ^3}{2 \kappa ^{3/2} \sqrt{\pi }} +\frac{\left(2
{\epsilon_\lambda}^2-2(2j+1){\epsilon_\lambda}+2j
(j+1)-9\right)\alpha^5}{16\kappa ^{5/2} \sqrt{\pi }} + \ldots\,
\ea
we resum $j$. Finally, one can find that
\ba I_{\lambda}(\alpha^2,\rho) &=& I^{(0)}_{\lambda}(\rho)
+I^{(1)}_{\lambda}(\rho) \rho \alpha^2 +\ldots \,, \ea where \ba
I^{(0)}_{\lambda}(\rho)&=&
 \frac{\rho ^5 }{8192}
\int_0^{\infty} dr
r^{11}
 K_{\epsilon_\lambda }\left(\frac{r}{\sqrt{2}}\right)
 K_{\epsilon_\lambda }\left(\frac{r \rho }{\sqrt{2}}\right)
\int_0^{\infty} d \kappa
 e^{-\frac{r^2}{8 \kappa }} \kappa^{-2}\,,
\ea
\ba
I^{(1)}_{\lambda}(\rho)&=&
\rho ^4
\int_0^{\infty} dr
r^{11}
\int_0^{\infty} d \kappa
\frac{e^{-\frac{r^2}{8 \kappa }}}{2097152
 \sqrt{2} \kappa ^5}
\nonumber \\[2mm] &&
 \left(16 r K_{{\epsilon_\lambda} +1}\left(\frac{r}{\sqrt{2}}\right)
 \left(2 ({\epsilon_\lambda} +1) K_{{\epsilon_\lambda} }\left(\frac{r \rho
 }{\sqrt{2}}\right)-\sqrt{2} r \rho K_{{\epsilon_\lambda}
 +1}\left(\frac{r \rho }{\sqrt{2}}\right)\right)
 \kappa ^2
\right.
\nonumber \\[2mm] &&
\left. +\, K_{{\epsilon_\lambda}
 }\left(\frac{r}{\sqrt{2}}\right) \left(32 ({\epsilon_\lambda}
 +1) r \rho K_{{\epsilon_\lambda} +1}\left(\frac{r \rho
 }{\sqrt{2}}\right) \kappa ^2
\right.
\right.
 \\[2mm] &&
\left. \left. +\sqrt{2} \left(r^4+16
 ({\epsilon_\lambda} -1) \kappa r^2 +8 \kappa ^2
 \left(\left(\rho ^2+1\right) r^2+8 ({\epsilon_\lambda} -7)
 {\epsilon_\lambda} -48\right)\right) K_{{\epsilon_\lambda} }\left(\frac{r
 \rho }{\sqrt{2}}\right)\right)\right)\,.  \nonumber \ea
Moreover we perform the integrals over $\kappa$, i.e.\
\ba I^{(0)}_{\lambda}(\rho)&=& \frac{ \rho ^5}{32} \int_0^{\infty}
dr r^9 K_{{\epsilon_\lambda}
 }\left(r\right) K_{{\epsilon_\lambda}
 }\left(r\rho \right)\,,
\ea
\ba
I^{(1)}_{\lambda}(\rho)&=&
\int_0^{\infty} dr
\frac{ \rho ^4 r^7}{128}
 \left( r K_{{\epsilon_\lambda} +1}\left(r\right)
\left(2 ({\epsilon_\lambda}
 +1) K_{{\epsilon_\lambda} }\left(r \rho \right)
-2 r \rho K_{{\epsilon_\lambda} +1}\left(r \rho \right)\right)
\right.
\nonumber \\[2mm] &&
\left.
+K_{{\epsilon_\lambda} }\left(r\right) \left(
 \left(\left(\rho ^2+1\right) r^2+4 ({\epsilon_\lambda} -3)
 {\epsilon_\lambda} -16\right) K_{{\epsilon_\lambda} }\left(r \rho \right)
\right.
\right.
\nonumber \\[2mm] &&
\left.
\left.
+2 ({\epsilon_\lambda} +1) r \rho
 K_{{\epsilon_\lambda} +1}\left(r \rho\right)\right)\right)\,.
\ea
and over $r$.
For $\epsilon_\lambda=1$
we get
\ba I^{(0)}_{T}(\rho)&=& \frac{192 \rho ^4 }{\left(\rho
^2-1\right)^8} \left(3 \rho ^8+178 \rho ^6+478 \rho ^4+178 \rho
^2+3\right) \nonumber \\[2mm] && -60 \frac{192 \rho ^6 \log (\rho^2)
}{\left(\rho ^2-1\right)^9}
 \left(\rho ^6+6 \rho ^4+6 \rho ^2+1\right) \,,
\lab{eq:I0T}
\ea
and
\ba I^{(1)}_{T}(\rho)&=& \frac{96 \rho ^3 }{\left(\rho
^2-1\right)^8} (\rho ^{10}+127 \rho ^{8}+712\rho ^6+712 \rho
^4+127\rho ^2+1) \nonumber \\[2mm] && -\frac{1152 \rho ^5 \log (\rho^2)
}{\left(\rho ^2-1\right)^9}
 \left(3 \rho ^8+33 \rho ^6+68\rho ^4+33 \rho ^2+3\right) \,,
\ea
while for $ \epsilon_\lambda=0$ the resulting expression looks
like
\ba I^{(0)}_{L}(\rho)&=& -\frac{1920 \rho ^5 }{ \left(\rho
^2-1\right)^8}
 \left(5 \rho ^6+37 \rho ^4+37 \rho ^2+5\right)
\nonumber \\[2mm] &&
+12 \frac{192 \rho ^5 \log (\rho^2 )}{\left(\rho ^2-1\right)^9}
\left(\rho^8+16 \rho^6 + 36 \rho^4 +16 \rho^2+ 1 \right)
\,,
\ea
\ba
I^{(1)}_{L}(\rho)&=&
-\frac{384 \rho ^4}{\left(\rho ^2-1\right)^8}
(7 \rho ^{8}+97 \rho ^6+212 \rho ^4+97 \rho ^2 +7)
\nonumber \\[2mm] &&
+\frac{576 \rho ^4 \log (\rho^2 )}{\left(\rho ^2-1\right)^9}
\left(\rho ^{10}+27 \rho ^8+112 \rho ^6+112 \rho ^4+27 \rho ^2+1\right)\,.
\lab{eq:I1L}
\ea
Moreover, the integral from eq.\ (\ref{eq:Atot}) reads \ba \hat
I_{\lambda}(\alpha^2=0,\rho)&\equiv& \hat I^{(0)}_{\lambda}(\rho)=
(\rho+\rho^{-1}) I^{(0)}_{\lambda}(\rho) - 2
I^{(1)}_{\lambda}(\rho) \,, \ea
so that
\ba \hat I^{(0)}_{T}(\rho)&=& \frac{384 \rho ^3
}{\left(\rho^2-1\right)^6} \left(\rho ^6+29 \rho ^4+29 \rho
^2+1\right) -\frac{4608 \rho ^5 \log \left(\rho
^2\right)}{\left(\rho^2-1\right)^7}
 \left(\rho ^4+3 \rho ^2+1\right)\,,
\ea
\ba
\hat I^{(0)}_{L}(\rho)&=&
-\frac{384 \rho ^4}{\left(\rho^2-1\right)^6}
\left(11 \rho ^4+38 \rho ^2+11\right)
+\frac{1152 \rho^4 \log \left(\rho ^2\right) }{\left(\rho ^2-1\right)^7}
\left(\rho ^6+9 \rho ^4+9 \rho ^2+1\right)\,.
\ea

\section{Variations of the action}
\label{sc:va}
The second variation of the action reads as
, i.e.\
\ba
 H^{(2)}&=&
 z_0^4
(d-5) (d-2)
h_{\alpha 0} h_{\alpha 0}
-\frac{1}{4} z_0^4
 ((d-7) d+20)
h_{\alpha \beta}
h_{\alpha \beta}
+
 \frac{1}{8} z_0^4
((d-7) d+16)
{\bar h}^2
\nonumber \\[2mm] & &
-\frac{1}{2} z_0^4 ((d-7) d+14)
h_{00} \bar h
+
 z_0^5 (d-6)
h_{\alpha 0} \partial_\alpha \bar h
-2 z_0^5 (d-4)
h_{\alpha 0} \partial_\beta h_{\alpha \beta}
\nonumber \\[2mm] &&
+ z_0^5 (d-7)
h_{\alpha \beta} \partial_0 h_{\alpha \beta}
-2 z_0^5 (d-3)
h_{\alpha \beta} \partial_\alpha h_{\beta 0}
-\frac{1}{2} z_0^5 (d-6)
\bar h \partial_0 \bar h
+ z_0^5 (d-4)
\bar h \partial_\alpha h_{\alpha 0}
\nonumber \\[2mm] &&
-\frac{3}{4} z_0^6
\partial_\alpha h_{\beta \gamma} \, \partial_\alpha h_{\beta \gamma}
+ \frac{1}{2} z_0^6
\partial_\alpha h_{\beta \gamma} \, \partial_\beta h_{\alpha \gamma}
+\frac{1}{4} z_0^6
\partial_\alpha \bar h \, \partial_\alpha \bar h
- z_0^6
h_{\alpha \beta} \partial_\alpha\partial_\beta \bar h
+2 z_0^6
h_{\beta \gamma} \partial_\alpha \partial_\beta h_{\alpha \gamma}
\nonumber \\[2mm] &&
- z_0^6
h_{\beta \gamma} \partial_\alpha \partial_\alpha h_{\beta \gamma}
- z_0^6
\partial_\alpha h_{\alpha \beta} \, \partial_\beta \bar h
+ z_0^6
\partial_\alpha h_{\alpha \gamma} \, \partial_\beta h_{\beta \gamma}
+\frac{1}{2} z_0^6
\bar h \partial_\alpha \partial_\alpha \bar h
-\frac{1}{2} z_0^6
\bar h \partial_\alpha \partial_\beta h_{\alpha \beta}
\ea
%


\end{document}